\documentclass[10pt,journal,compsoc]{IEEEtran}
\usepackage[sort,nocompress]{cite}
\usepackage[pdftex]{graphicx}
\usepackage{amsmath}
\usepackage{amssymb}

\usepackage{algorithmic}
\usepackage{array}
\usepackage[caption=false,font=footnotesize,labelfont=sf,textfont=sf]{subfig}
\usepackage[font=small,labelformat=original, justification=centering]{caption}
\usepackage{url}
\usepackage{paralist}
\usepackage{comment}
\usepackage[ruled, linesnumbered,noend]{algorithm2e}
\usepackage{color}
\usepackage{soul}

\newcolumntype{L}{>{\centering\arraybackslash}m{1.68cm}}
\newcolumntype{D}{>{\centering\arraybackslash}m{3cm}}
\newcolumntype{R}{>{\centering\arraybackslash}m{2.9cm}}
\newcolumntype{A}{>{\centering\arraybackslash}m{2.3cm}}
\newcolumntype{B}{>{\centering\arraybackslash}m{1.65cm}}
\newcolumntype{E}{>{\centering\arraybackslash}m{2.2cm}}
\newcolumntype{F}{>{\centering\arraybackslash}m{1.7cm}}
\newcolumntype{G}{>{\centering\arraybackslash}m{19 pt}}
\newcolumntype{J}{>{\centering\arraybackslash}m{20 pt}}
\newcolumntype{H}{>{\centering\arraybackslash}m{32 pt}}
\newcolumntype{I}{>{\centering\arraybackslash}m{26 pt}}
\newcommand*{\myfont}{\fontfamily{pcr}\selectfont}
\DeclareTextFontCommand{\myemph}{\myfont}
\newlength{\textfloatsepsave} 
\newcommand\fullwidt{0.59}
\newcommand\filterwidt{0.270}
\newcommand\revMark[1]{#1}
\begin{document}
%
\title{Understanding Bulk-Bitwise Processing In-Memory Through Database Analytics}
%
%
%
%

\author{Ben~Perach,
        Ronny~Ronen,~\IEEEmembership{Fellow,~IEEE,}
        Benny~Kimelfeld, 
        and~Shahar~Kvatinsky,~\IEEEmembership{Senior Member,~IEEE}
\IEEEcompsocitemizethanks{\IEEEcompsocthanksitem B. Perach, R. Ronen, and S. Kvatinsky are with the Andrew and Erna Viterbi Faculty of Electrical and Computer Engineering, Technion -- Israel Institute of Technology, Haifa.\protect\\
E-mail: benperach@campus.technion.ac.il, ronny.ronen@technion.ac.il, shahar@ee.technion.ac.il.
\IEEEcompsocthanksitem B. Kimelfeld is with the Taub Faculty of Computer Science, Technion -- Israel Institute of Technology, Haifa.\protect\\
E-mail: bennyk@cs.technion.as.il.}
\thanks{Manuscript received September 30, 2022. 
This work was supported by the European Research Council through the European Union's Horizon 2020 Research and Innovation Programme under Grant 757259.}}

%

\markboth{IEEE Transactions on Emerging Topics in Computing,~Vol.~X, No.~X, Month~YYYY}%
{Perach \MakeLowercase{\textit{et al.}}: Understanding Bulk-Bitwise Processing In-Memory Through Database Analytics}



\IEEEtitleabstractindextext{%
\begin{abstract}
Bulk-bitwise processing-in-memory (PIM), where large bitwise operations are performed in parallel by the memory array itself, is an emerging form of computation with the potential to mitigate the memory wall problem. This paper examines the capabilities of bulk-bitwise PIM by constructing PIMDB, a fully-digital system based on memristive stateful logic, utilizing and focusing on in-memory bulk-bitwise operations, designed to accelerate a real-life workload: analytical processing of relational databases. 
We introduce a host processor programming model to support bulk-bitwise PIM in virtual memory, develop techniques to efficiently perform in-memory filtering and aggregation operations, and adapt the application data set into the memory.
To understand bulk-bitwise PIM, we compare it to an equivalent in-memory database on the same host system. We show that bulk-bitwise PIM substantially lowers the number of required memory read operations, thus accelerating TPC-H filter operations by 1.6$\times$--18$\times$ and full queries by 56$\times$--608$\times$, while reducing the energy consumption by 1.7$\times$--18.6$\times$ and 0.81$\times$--12$\times$ for these benchmarks, respectively. Our extensive evaluation uses the gem5 full-system simulation environment. The simulations also evaluate cell endurance, showing that the required endurance is within the range of existing endurance of RRAM devices.    
\end{abstract}

}

\maketitle

\IEEEdisplaynontitleabstractindextext

%
\IEEEpeerreviewmaketitle

\IEEEraisesectionheading{\section{Introduction}\label{sec:introduction}}

%
%
%

\IEEEPARstart{A}{} promising technique for accelerating memory-bounded computation moves some computation from the processor to the memory unit (\textit{processing-in-memory}, PIM), thus reducing the data movement latency and energy.
Due to the ever-increasing performance and energy gaps between processing units and memory, numerous PIM architecture types and techniques, such as near-memory~\cite{Kepe2019}, analog~\cite{ISAAC}, and CAM~\cite{NVQuery,CATCAM}, have gained attention in recent years.
One particular PIM technique relies on using the memory cell arrays to both store the data and perform logical operations. This PIM technique can be implemented with DRAM~\cite{AMBIT,SIMDRAM} or nonvolatile memory technologies~\cite{MOUSE,CONCEPT, FloatPIM,RAPID,Lu2021,Pinatubo}. Because of the regular structure of the memory cell array, logical operations can be performed concurrently between many sets of cells, resulting in bulk-bitwise operations. Hence, in addition to the reduced data movement compared to other PIM techniques~\cite{Reuben17}, bulk-bitwise PIM can achieve significant operation throughput, potentially being performed on all memory cell arrays in parallel. 
Nevertheless, building such a system and executing complete applications using bulk-bitwise PIM remains a challenge. 

Previous works focused on certain aspects of bulk-bitwise PIM. These aspects include exploration of technology~\cite{AMBIT,Pinatubo,Giannopoulos2020,RACER,MOUSE}, programming language abstraction~\cite{SIMDRAM,Zhou2021}, memory interface~\cite{CONCEPT}, logic design and abstraction~\cite{SIMPLER,Lu2021,Talati2016}, and combining bulk-bitwise with other PIM technologies~\cite{NVQuery,FloatPIM}.
A full system perspective of bulk-bitwise PIM is, however, still missing. Insights about the benefits and limitations of such an approach need to be educed. Aspects such as what applications are suitable for this type of processing and what the system requirements have yet to be resolved.
In this paper, our goal is to investigate bulk-bitwise PIM and determine how it can be utilized. We ask the above questions and seek to understand the performance gains, drawbacks, and bottlenecks of bulk-bitwise PIM.

To achieve our goal of bulk-bitwise PIM investigation, we propose PIMDB, an example of a full PIM system based on bulk-bitwise logic. PIMDB accelerates operations from a real-world application: analytical processing of relational databases. \revMark{We focus on a single application case study to concisely present its adaptation for bulk-bitwise PIM (for software and hardware) and its performance analysis}. By utilizing only bulk-bitwise PIM techniques and addressing the requirements of a real-world application, the benefits and limitations of bulk-bitwise PIM can be identified. PIMDB is also the first to introduce a host programming model that works in virtual memory for PIM bulk-bitwise operations. The programming model is an essential part of our evaluation, as it reflects the interaction of the software, host hardware, and PIM hardware.  

We target analytical processing of relational databases as our example application for several reasons. First, it is an important real-world application, found in business decision support processes~\cite{Hasso2013,TPCH}. Second, relational databases are stored in tabular structures called \textit{relations}, each a collection of small and independent items called \textit{records}. Records can be stored in a small physical memory region and operated on concurrently. The ability to fit each independent data unit into a small physical memory region is crucial, as bulk-bitwise operations require the operands from the same data unit to be in physical proximity in the memory cell array~\cite{SIMDRAM,Talati2016}. Third, a database can reside in memory in a dedicated structure and be repeatedly used, amortizing the one-time cost of loading the database into the PIM module. When executing queries, no data transfer to the PIM module is required before execution starts. Fourth, database analytical processing is a memory-bound application~\cite{Hasso2013}, requiring the performance of simple operations (\emph{e.g.}, comparison, addition, multiplication) on large data sets (from a few to thousands of GBs~\cite{TPCH}). Fifth, analytical processing queries perform aggregation (\textit{e.g.}, sum, average) on many database records~\cite{Hasso2013}.
Performing the aggregation, or part of the aggregation, in memory can potentially \textbf{reduce data movement by several orders of magnitude}. 

PIMDB uses a byte-addressable, nonvolatile memristive memory utilizing stateful logic~\cite{Borghetti2010,Reuben17, MOUSE,CONCEPT,FloatPIM,RAPID,Lu2021}, which stores a copy of the database. We chose memristive stateful logic as the underlying bulk-bitwise technology as it can operate on both bitlines and wordlines~\cite{Talati2016}, enabling a richer functionality compared to other (\emph{e.g.}, DRAM-based) bulk-bitwise PIM. Nevertheless, our techniques are not limited only to memristive stateful logic; they can be applied to other bulk-bitwise PIM technologies as well.
PIMDB's speedup is primarily driven by the reduction of memory accesses, where the PIM operations reduce the amount of data read out of the memory array. In this paper, we show that over 99\% of the reads can be eliminated for some queries when using bulk-bitwise operations.
The proposed design supports in-memory record filtering and aggregation operations over database relations (similarly to commercial in-storage analytical processing~\cite{FlexPushdownDB}). For queries involving a single relation, most query execution can be performed solely in PIMDB.
For queries involving multiple relations, PIMDB performs the filtering while the rest of the query is executed at the host in a standard non-PIM fashion.

PIMDB is evaluated in a gem5 full-system simulation~\cite{gem5}, using queries from the TPC-H benchmark~\cite{TPCH} where the SQL queries are compiled using our in-house compiler. The proposed design uses stateful logic with memristive devices~\cite{Reuben17}, specifically, RRAM-based MAGIC NOR gates~\cite{MAGIC}. As PIMDB focuses on bulk-bitwise operations, seeking to reveal their strengths and weaknesses, our evaluation is of use for other PIM architectures beyond the context of databases.

In summary, this paper makes the following contributions:
\begin{compactitem}
\item We evaluate the performance of and derive insights regarding bulk-bitwise PIM operations in general and in stateful logic specifically.
\item We define a programming model for bulk-bitwise logic-based PIM architectures, enabling the use of bulk-bitwise logic as part of host applications in virtual memory space.
\item We present PIMDB, a PIM-based accelerator for analytical processing of relational databases.
\item We establish database querying as an empirical proof of application for \revMark{bulk-bitwise} PIM where the latter accelerates the performance of the underlying operations.
\item We demonstrate speedup improvement of 1.6$\times$--18$\times$ on filter operations and 56$\times$--608$\times$ for full queries from TPC-H, and an energy reduction of 1.7$\times$--18.6$\times$ and 0.81$\times$--12$\times$, respectively.
\item We provide a gem5 full-system simulation for bulk-bitwise PIM\footnote{https://github.com/benperach/gem5\_PIM\_extension}.
\end{compactitem}

\section{Background}

\subsection{Bulk-Bitwise PIM and Stateful Logic}
\label{subsec:stateful_logic}

Bulk-bitwise PIM refers to a class of PIM techniques categorized by the location and capabilities of the PIM processing elements. In bulk-bitwise PIM, the computation elements are the memory cells and their periphery circuits (\textit{e.g.}, sense amplifiers, voltage drivers, decoders), thus the memory storage medium itself can compute (\textit{i.e.,} PIM, referred to also as \textit{processing using memory}). Several technologies were suggested for implementing such processing capabilities, including DRAM~\cite{AMBIT,SIMDRAM} and emerging nonvolatile memory technologies~\cite{MOUSE,CONCEPT,FloatPIM,RAPID}, all of which execute simple logic operations (\textit{e.g.}, AND, OR, NOT). These simple logical operations use the memory cells as inputs and write the result directly to a memory cell. Furthermore, due to the regular structure of memory cells in a memory array, these operations can be concurrently performed on many sets of cells within an array, \textit{i.e.}, executing bitwise operations. As different memory arrays are independent circuits, many memory arrays can concurrently perform these bitwise operations, resulting in a wide bitwise operation, \textit{i.e.}, a \textit{bulk-bitwise} operation. The concurrent operation on many memory arrays, each performed on many sets of cells, produces substantial computational throughput.

Implementing bulk-bitwise PIM using emerging memory technologies is often referred to as stateful logic~\cite{Borghetti2010,MAGIC} and specifically refers to memristive technology (\textit{e.g.}, ReRAM, PCM, MTJ)~\cite{Zahoor2020}. 
Memristive devices are devices that can change their resistance when electrically stimulated.

\setlength{\textfloatsep}{0.1cm}
\setlength{\floatsep}{0.1cm}
\begin{figure}
\centering%
\subfloat[\label{fig:crossbar_mat}]{\includegraphics[width=0.40\linewidth]{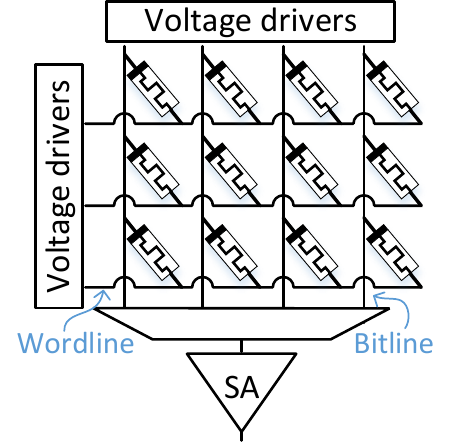}}\hfill%
\subfloat[\label{fig:crossbar_magic}]{\includegraphics[width=0.52\linewidth]{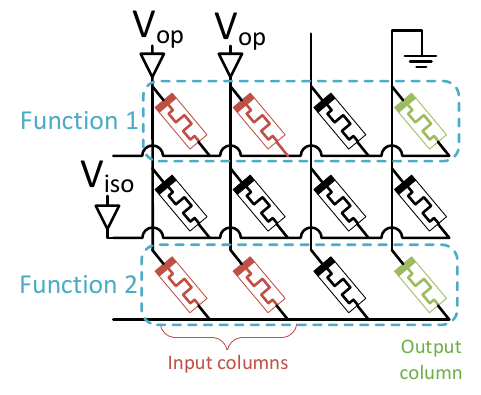}}%
\caption{(a) A $3\times4$ memristive crossbar array with a single device per cell (1R~\cite{Xu2015}) and a single sense amplifier. (b) Logic operations between crossbar columns. The inputs are marked in red, the output in green. Devices in black do not participate in the logic operation. Specifically, row 2 is isolated and does not participate in the operation. Row-wise operations can be performed similarly~\cite{Talati2016}.}%
\label{fig:crossbar}%
\end{figure}

The resistive state of memristive devices is nonvolatile, and can, therefore, be used to construct nonvolatile memories~\cite{Xu2015}, where each memory cell is a single memristive device. 
A memory array of memristive devices is constructed as a crossbar~\cite{Xu2015} (Fig.~\ref{fig:crossbar_mat} shows an example of a $3\times4$ crossbar array),
where the horizontal and vertical wires are referred to as wordlines and bitlines, respectively. 
These memories are random-access memories, as they can read or write specifically selected cells.

Bulk-bitwise operations are performed by applying a voltage across the bitlines and/or wordlines in a certain manner, independently of the stored cell values. Furthermore, this logic technique can be applied in parallel on a crossbar row or column~\cite{Talati2016}, as illustrated for column-wise operations in Fig.~\ref{fig:crossbar_magic}. It is possible to exclude certain rows or columns from a parallel operation. For example, Fig.~\ref{fig:crossbar_magic} shows how a row can be excluded in a crossbar with a single memristive device per cell (1R crossbar~\cite{Xu2015}). The row exclusion from a parallel operation is done by setting the corresponding wordline to an isolation voltage, leaving the output device undisturbed~\cite{Talati2016}. Note that the input and output operands for stateful logic, as for other bulk-bitwise technologies, must reside on the same wordline/bitline. This physical proximity of operands poses a challenge for bulk-bitwise PIM, as the data has to be arranged in an appropriate way to enable PIM use.

An example of a stateful logic technique is MAGIC NOR~\cite{MAGIC,FloatPIM}. Since NOR is functionally complete, any logical function can be performed using a series of NOR operations (requiring additional devices for the intermediate values). In PIMDB, we use MAGIC NOR as the basic stateful logic gate. Nevertheless, most of the concepts presented for MAGIC NOR apply to any functionally complete bulk-bitwise PIM technology~\cite{AMBIT,Pinatubo,SIMDRAM} (all concepts apply as long as row-wise and column-wise bulk-bitwise operations are available).

Memristive random-access memory can be used as the system's main memory. Memristive memory has several potential advantages over DRAM; the small device size and the dense layout of memristive arrays result in high-capacity memory~\cite{Talati2016}. The nonvolatility of memristive devices eliminates the need for memory refreshes, reducing idle power consumption. The disadvantages of memristive memory are its longer access time relative to DRAM and its limited cell endurance~\cite{Zahoor2020}. Due to these limitations, to date, memristive memories have not been replacing DRAM as main memory; rather, memristive memories have been augmenting DRAM as a new memory hierarchy tier~\cite{OptaneDC}.

\subsection{Memory Organization}
\label{subsec:mem_org}
DRAM and memristive memory are constructed similarly. The memory is divided into \emph{channels, ranks, chips}, and \emph{banks}~\cite{Xu2015,MemSys}. Each channel has a dedicated bus, and though the ranks in a channel share the bus, they can operate independently. Each rank comprises one or more banks and one or more chips. Each bank is logically distributed across several chips 
where all bank parts operate in lockstep. The banks of a rank share the rank's memory bus interface and can operate independently.

For practical reasons, banks are not built as a single memory array in each chip; they are split into units called \emph{subarrays}~\cite{Xu2015}. In memristive memory, a subarray comprises several crossbars (including periphery circuits, \emph{e.g.}, sense amplifiers, voltage drivers, decoders), which operate in lockstep. When accessing a bank as part of a traditional memory read/write operation, only the necessary subarray is activated.

\begin{figure}
\centering
  \includegraphics[width = 0.82\linewidth]{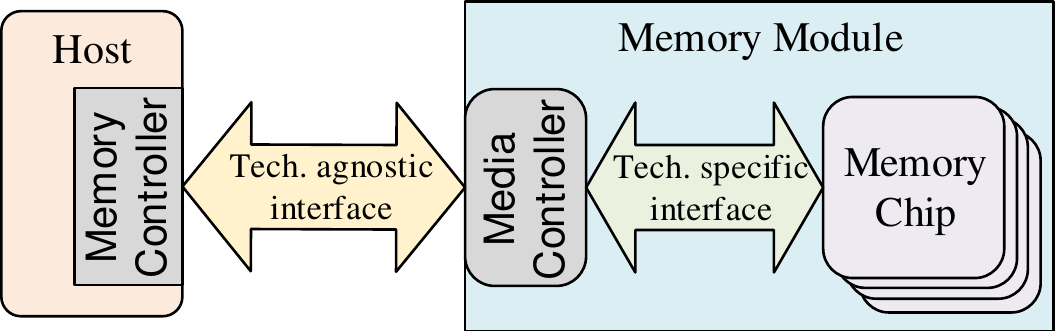}
\caption{The general structure for emerging interfaces (\textit{e.g.}, OpenCAPI, GenZ, CXL). The host-memory interface is agnostic to the memory technology; the technology-specific operations are performed by the memory module.} 
\label{fig:media}
\end{figure}

Traditionally, the host--memory interface is designed especially for DRAM (\textit{e.g.,} the DDRx standard), which results in lower performance for memristive technologies~\cite{CONCEPT}. Furthermore, traditional memory interfaces specify a fixed set of memory operations (\emph{e.g.}, read, write, refresh, \textit{etc.}), limiting the use of novel memory designs (such as PIM-enabled memory). To address these issues, new technology-agnostic interfaces have been proposed, \emph{e.g.}, OpenCAPI
, GenZ
, and CXL
. These new interfaces issue technology-independent operations, with possible vendor extensions, and delegate the technology-dependent management to the memory module. The general structure of these new interfaces is shown in Fig.~\ref{fig:media}, where the host--side memory controller communicates with a \emph{media controller} using a technology-agnostic interface, while the media controller communicates with the memory chips using a technology-specific interface (\textit{e.g.} DDRx for DRAM, R-DDR~\cite{CONCEPT} for RRAM). The media controller is part of the memory module, enabling optimization of the media controller for the specific memory technology and design.

\subsection{Relational Databases and Analytical Processing}
A relational database is a data model that organizes data into relations (tables)~\cite{Hasso2013}. A relation can be thought of as a set of records and a set of attributes. Attributes may be of different types, and each record is assigned a value for each attribute. In the relation, each record is represented by a different row, and each attribute is represented by a different column. The value of a record for a specific attribute is the relation entry located at the intersection of the record row and the attribute column. 
Fig.~\ref{fig:table_layout_relation} visualizes a relation.

A query over the database is a question about the content of the database's relations, \emph{i.e.,} about attributes of a specific record or a group of records fulfilling a certain condition. 
Analytical processing refers to a class of queries requiring the selection of records according to some criteria and their aggregation (\textit{e.g.}, sum, average)~\cite{Hasso2013}. These kinds of queries appear in applications such as business decision support processes and have dedicated benchmarks~\cite{TPCH}.

Databases can be represented and stored in various data structures. Two main categories of such data structures are \emph{row-store} and \emph{column-store}. In row-store, a relation is stored as an array of records, each record having a field for each attribute. In column-store, each relation attribute is stored as a single array, and a relation is a collection of such arrays.  
Due to data locality, row-store is better suited for accessing single records, while column-store is better suited for accessing several attributes over many records~\cite{Hasso2013}. 
In PIMDB, reads and writes access relations as row-stored, while PIM operations, due to their access pattern, access the same relations as column-stored (see Section~\ref{sec:db_with_mMPU}).

\section{PIMDB Architecture Overview}
\label{sec:arch}
We now present the proposed PIMDB design. This section also presents the PIMDB programming model, which specifies the PIM interface for the software (\emph{e.g.}, for compiler or library writers; for databases, these are the database management system writers). The programming model can be used by a compiler from a high-level language such as SQL, as used for our evaluation (Section~\ref{subsec:compile}).

The PIMDB system, shown in Fig.~\ref{fig:arch}, comprises a host processor, DRAM main memory, and additional byte-addressable memory ranks based on RRAM to support bulk-bitwise MAGIC NOR stateful logic operations~\cite{MAGIC}. These bulk-bitwise PIM ranks are termed \textit{PIM modules}, and their memory is mapped to the host memory address space and accessed using standard reads and writes. The PIM modules have a predefined instruction set, performing operations, such as add, multiply, less-than, \emph{etc.}

\subsection{Programming Model}
\label{subse:programmin_model}

The programming model defines how the host software interacts with the PIM modules, \textit{i.e.}, how it utilizes the PIM module's instructions on the relevant data. We show here how this interaction can support virtual memory and how the software can control the physical proximity of PIM operands.

To operate on data in a PIM module, the host software sends a sequence of requests\revMark{, named \emph{PIM requests},} to the PIM module. \revMark{PIM requests} contain address and data fields, similar to a memory write operation. In PIMDB, for example, PIM requests are sent by a dedicated host instruction, similar to loads and stores. Regardless of how PIM requests are sent, each PIM request contains information for a single PIM instruction, \emph{e.g.}, the instruction's opcode and locations of operands. The possible instructions are determined by the PIM module's ISA, possibly high-level operations such as addition and multiplication (the translation to basic operations supported by the PIM module's crossbars, \emph{e.g.}, a sequence of MAGIC NOR operations as in PIMDB, is done at the PIM module).

\textbf{Virtual memory}: Since bulk-bitwise PIM is done in the main memory of the host processor, and since user-level processes use virtual memory, it is essential for bulk-bitwise PIM to support virtual memory to be integrated into such a system. Furthermore, supporting virtual memory is essential not only from a compatibility point of view but because of the advantages virtual memory provides: holding multiple and independently operated-on data structures (such as database relations) in memory and enabling dynamic data allocation. 
Virtual memory support can be achieved by allowing PIM instructions to operate within the boundaries of a specific virtual memory page, \textit{i.e.}, a PIM request is sent to an individual virtual page and can only access and change data within that page. Such a mechanism utilizes the existing virtual memory implementation, issuing the PIM requests with virtual addresses that are translated in the usual way \revMark{to physical addresses}, and thereafter sent to the required memory module.  
When operating on a data structure spanning multiple pages, a PIM request should be sent to each page belonging to that data structure.

Nevertheless, there are challenges in supporting virtual memory as suggested: (1) Bulk-bitwise PIM operation's operands must be located on the same memory crossbar, possibly on the same crossbar row or column~\cite{AMBIT,CONCEPT}. The physical memory abstraction, where the memory channels, ranks, banks, and crossbars appear as a monolithic address range, hides the knowledge required for such data allocation. Moreover, the virtual memory abstraction hides the specific physical memory address from the user-level software. To allocate data for bulk-bitwise PIM, these two abstraction levels, virtual and physical memory, must be circumvented in some manner. (2) An essential property for bulk-bitwise PIM performance is its high parallelism. Hence, a PIM request should initiate its specified PIM instruction on as many crossbars as possible. Restricting a single PIM request to a single page, however, will restrict the number of crossbars the PIM request targets, limiting the benefit of bulk-bitwise PIM. This situation is especially acute when dealing with standard 4KB pages, which are smaller than a single crossbar~\cite{Xu2015,MemSys}. 
In the following paragraphs, we address these challenges, explaining how our programming model entails the address mapping of physical addresses to memory cells and utilizes huge-pages. 

\textbf{Using huge-pages:} To maximize the number of crossbars within a page and increase the parallelism of a single bulk-bitwise operation, we use huge-pages. Huge pages can range in size from a few MB to a few GB (PIMDB uses 1GB huge-pages), sufficiently large to contain multiple memory crossbars.
\revMark{ These huge pages are only required for the PIM memory, which is in addition to the standard DRAM main memory (see the start of Section~{\ref{sec:arch}} and Figure~{\ref{fig:arch}}). Hence, using huge pages for PIM does not affect the usability and performance of the DRAM main memory.}

\textbf{PIM operand allocation:}
To address the challenge of how to allocate PIM operands, we give the software fine-grained control over the location of values in the memory array. Using this fine-grained control, the software will decide how each data structure will be located across crossbars, crossbar rows, and crossbar columns. As the software is the entity that knows how the data is going to be operated on, allowing it the flexibility to set data structures according to its specific needs will allow our programming model to suit different applications. To enable such fine-grained control, the virtual-to-physical address translation and the physical-address to memory cell translation have to be controlled in some manner by the software.

To control the virtual-to-physical address translation we note that the address translation does not change the page offset. This allows the software to directly control the page offset bits of the physical address. Because our programming model allows bulk-bitwise PIM operations to be performed only within a single page, controlling the page offset, which encodes the relative location within the page, is sufficient for PIM data allocation.

\begin{figure}
\centering
\includegraphics[width=0.80\linewidth]{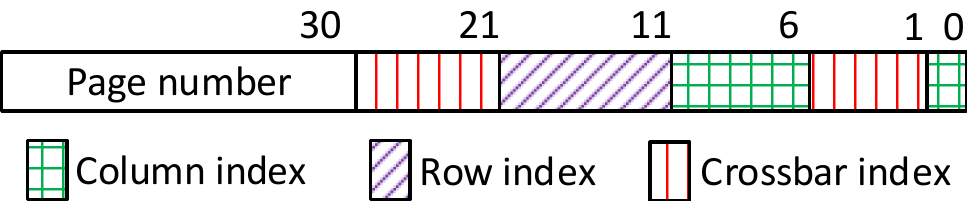}%
\caption{Example of address mapping, revealed to the programmer as part of the programming model. PIMDB uses this configuration for 1GB pages with $1024\times 512$ cell crossbars (see Section~\ref{subsec:PIM_Module_Evaluated_Details}). The fields of this mapping are not consecutive due to the internal structure of the memory, \emph{e.g.}, the number of accessed crossbars in a read/write and the number of bits read/written together from a single crossbar.}
\label{fig:address_map}
\end{figure}

To control the physical-address to memory cell translation, our programming model reveals the necessary details of that translation to the software. The necessary details are which bits in the physical page offset decode the crossbar index within the page, and the row and column index within the crossbar. An example of such mapping is shown in Fig.~\ref{fig:address_map}. Hence, by controlling the crossbar, row, and column bits in a virtual address, user-level software can target and manage each cell in each crossbar of a single page. By using loads, stores, or PIM requests, the value of specific cells can be read, written, or operated on. Note that a continuous data unit on a crossbar row (\emph{e.g.}, the relation record in Section~\ref{subsec:db_layout}) is not necessarily continuous in the virtual address space. Non-continuous data structures can be handled by software abstraction (as our compiler does in Section~\ref{subsec:compile}).

\textbf{PIM requests:} \revMark{PIM requests encode the details of the PIM operation they transfer with their address and data.} The page number in the \emph{address} specifies the page for which the instruction is meant. PIM requests target all crossbars within their targeted page, thus ignoring the crossbar index field within the address. The column and/or row index fields within the page offset indicate the location of the instruction \emph{result} within each crossbar. The PIM request's \emph{data} specify all additional fields: opcode, location of input operands within the crossbar, immediate operands, \emph{etc.} As this work focuses only on bulk-bitwise operations, no PIM instructions are allowed to move data between crossbars or operate on operands from different crossbars, whether within the same page or among pages. To move data among crossbars, standard reads and writes should be used. 

For memory consistency purposes, the same rules that apply to write requests apply to PIM requests, as PIM requests are similar to writes in the sense that they change the memory data. 
To enforce the desired memory ordering of reads, writes, and PIM requests, programmers should use the memory order rules of the host processor, treating PIM requests as write requests, as well as using the available ordering primitives (\emph{e.g.}, memory fences).  
Furthermore, the processor caches ignore PIM requests and forward them to the memory. It is the responsibility of the programmer to ensure coherency between the host caches and the PIM memory, using cache flushes. 

The use of PIM requests abstract the PIM module details for the host hardware. The host hardware is simply required to transfer the PIM requests to the PIM modules -- the host is not required to know their exact meaning. Therefore, the host is able to work with different PIM modules, utilizing different bulk-bitwise PIM technologies and instruction sets. The software determines the address and data of the PIM requests, cognizant of their full meaning, and thus controls the PIM modules. Hence, our proposed programming model is general and can apply to applications other than database analytical processing. The extent of the host hardware knowledge is presented in Section~\ref{subsec:host_modifications}.

\textbf{Additional computation area:}
To execute PIM instructions, the PIM modules require some crossbar area for their internal computations (for storing intermediate results). The software is responsible for allocating this crossbar area by configuring the PIM module. To provide the software with the necessary knowledge, the computation area requirements for each PIM instruction are given as part of the PIM module instruction set architecture. Each page has its own computation area configuration, which applies to all crossbars within the page. This configuration, conveyed via a special PIM instruction to the targeted page, configures the page but does not access the crossbars themselves. The software can change the page configuration at any time to suit the needs of the upcoming PIM instructions.

\subsection{The PIM Module}
\label{subsec:pim_module}
\begin{figure}
\centering
\includegraphics[width = 0.9\linewidth]{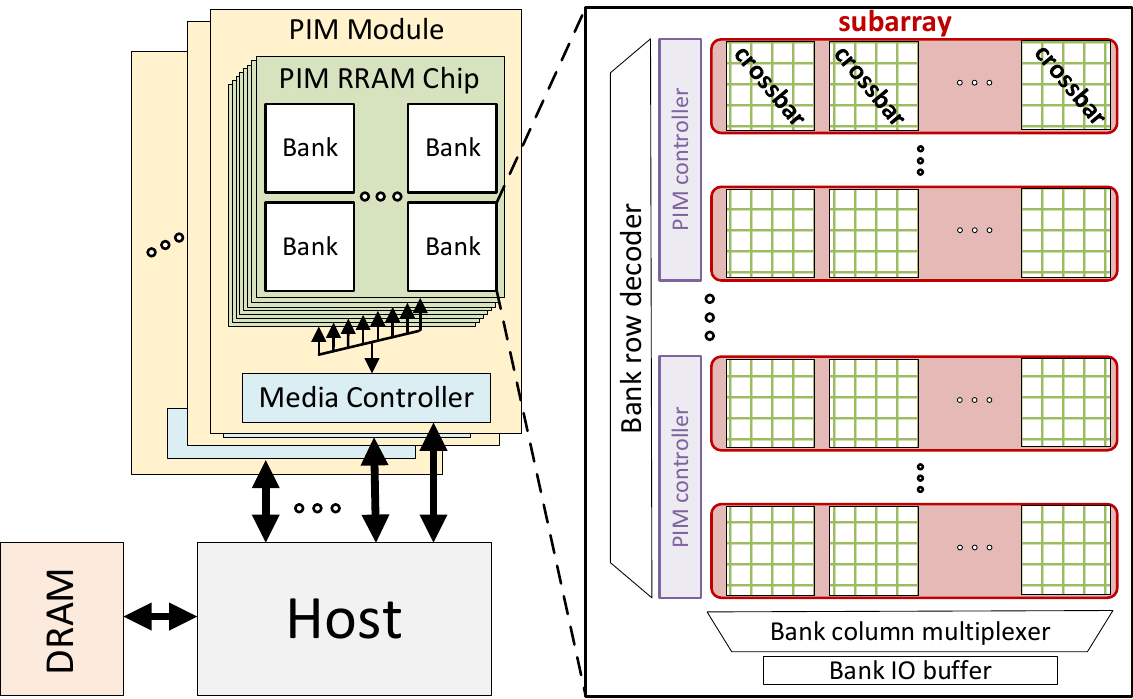}
\caption{PIMDB system structure and PIM-enabled memory bank schematic.}%
\label{fig:arch}
\end{figure}

The proposed PIM module is a single memory rank and is constructed similarly to an RRAM memory module: several RRAM memory chips and a single media controller chip (as shown in Fig.~\ref{fig:arch}). The media controller communicates with the host memory controller using the OpenCAPI interface~\cite{OpenCAPI}, and with the memory chips through an interface similar to the R-DDR interface~\cite{CONCEPT}. The media controller schedules the incoming requests using a First-Ready-First-Come-First-Serve (FR-FCFS) policy, which considers the dependencies between reads, writes, and PIM requests.

The R-DDR interface is similar to DDRx~\cite{MemSys}. It has an address bus, data bus, and several command signals. The address bus and command signals are connected in parallel to all memory chips, making the chips work in lockstep, while the data bus is divided among the chips. The difference between R-DDR and DDRx is in the operations they specify; R-DDR targets RRAM technology, while DDRx targets DRAM (see~\cite{CONCEPT} for details). For the interface used in this work, reads and writes are kept the same as in R-DDR, while PIM operations are handled differently. A PIM operation is specified, similar to a read or write operation, by a certain combination of the command signals. On issuing a PIM request to the RRAM chips, the media controller sets the command signals accordingly, passes the address of the PIM request to the address bus, and duplicates the PIM request's data to all the chip data buses (according to the timing required by the interface), issuing the same PIM request to all chips in parallel.

Each memory chip is constructed similarly to an RRAM memory chip~\cite{Xu2015}. Reads and writes are performed as in a standard RRAM chip. To support PIM operations, \emph{PIM controllers} are incorporated in the memory banks, as illustrated in Fig.~\ref{fig:arch}. Each PIM controller controls all crossbars across multiple subarrays, all of which belong to the same memory space of a single huge-page. A huge-page is assigned to a single bank of a single memory rank, requiring several PIM controllers in each chip to cover all the page's subarrays.
When a PIM request is issued to the chip, its information is routed to the destination bank as a read or a write, and then intercepted by the targeted PIM controllers. Each targeted PIM controller then issues the required MAGIC NOR sequence of operations to all crossbars connected to it, according to the specific instruction.
While a PIM controller is executing its operation, since it does not require the use of the bank's global resources, the bank can perform reads, writes, or PIM operations on subarrays that are not being controlled by that PIM controller.

The media controller is responsible for orchestrating the requests sent to the memory chips and for identifying the bank or subarray that can operate (similar to the operation of the memory controller in a DDRx interface).

PIM architectures occasionally need inter-crossbar data movements. These data movements can be accomplished indirectly via simple, yet slower, CPU load/store instructions or using faster, but more complex, direct inter-crossbar communication~\cite{RACER,FloatPIM}. Direct inter-crossbar communication can be added to PIMDB at the cost of added hardware and complexity. Here, we decided not to include direct inter-crossbar communication, as adding such a PIM technique obscures the explicit contribution of bulk-bitwise PIM and our evaluation (Section~\ref{sec:results}) shows that PIMDB achieves significant speedup without it. 
Hence, inter-crossbar data movements are performed via standard load/stores. 

\subsection{Instruction Design}
\label{subsec:InstructionImpelemntation}
PIM controllers perform each PIM instruction \revMark{as a} sequence of bulk-bitwise NOR~\cite{MAGIC} operations. This sequence is implemented as a finite-state-machine (FSM) included in each PIM controller. Some instructions require operations on in-memory values of various lengths (Section~\ref{subsec:PIMop}). Implementing a predefined operation sequence for every possible length, such as suggested in~\cite{SIMPLER,SAID}, might require numerous FSM states. Instead, we implement an $n$-bit operation by iterating a single-bit operation $n$-times. 
The operand's length determines the number of iterations.
For instance, addition and multiplication with variable-length operands are implemented by iterating over a full-adder sequence. Such implementation increases the length of the bulk-bitwise logic sequence for each instruction, since there is no optimization across iterations, but it supports many operand lengths with the same states in the FSM. Moreover, such implementation reduces the number of cells required for the internal computation, given that the iterated short logic sequence can naturally reuse the same cells.

Additionally, we added an optimization that uses immediate values (instruction constants) in the control paths. When performing PIM operations with an immediate value, the control sequence of the FSM depends on this value. Thus, instead of writing the immediate value to the crossbar (possibly duplicating it on every row) 
and performing the operation between two in-memory values, the bulk-bitwise logic sequence is determined according to the immediate value. An example of an equality comparison between an in-memory value and an immediate value is shown in Algorithm~\ref{algo:imm_eq}. Note that an instruction uses the same FSM states for all immediate values. The use of the immediate value for control reduces the number of logic operations and the latency, eliminates the need to store the immediate value in each row, and improves the lifetime of the PIM module due to fewer cell writes. To the best of our knowledge, this work is the first to implement bulk-bitwise logic operations in this manner.
 
\subsection{Modifications to the Host Processor}
\label{subsec:host_modifications}
To support PIM requests, some adjustments to the host are needed. First, the host must be capable of issuing a software-initiated PIM request, similar to the way the host sends a read/write request. For memory operation ordering, a PIM request is considered as if it were a write request. Caches at the host, however, should not use the content of the PIM requests' address or data since their meaning is dependent on the PIM specifications. In this work, caches forward PIM requests as is and do not change their state or data when encountering PIM requests. As explained in Section~\ref{subse:programmin_model}, it is the programmer's responsibility, \emph{e.g.}, using flushes, to ensure coherence between the memory and the caches when using PIM requests. We leave further investigation of PIM request consistency and coherence support for future work.

Further, the host processor should have memory controllers that support OpenCAPI. These controllers are assumed to preserve the order of requests passed to them since they cannot know the meaning of the PIM requests' data and addresses.

\begin{algorithm}[t]
\small
\SetAlgoNoLine
\DontPrintSemicolon
\SetKwInOut{Input}{Input}\SetKwInOut{Output}{Output}
\Input{$v_{n-1}...v_0$ - in-memory value bits\\$c_{n-1}...c_0$ - immediate bits at the PIM controller\\}
\Output{in-memory bit $m_{eq}$}
\KwResult{$m_{eq}=1$ if for all $i\ \ v_i=c_i$, else $m_{eq}=0$}
 \BlankLine
 $InMemory(m_{eq}\leftarrow1)$\;
 \ForEach{$i\in[0,..,n-1]$}{
  \eIf{$c_i = 1$}{$InMemory(m_{eq}\leftarrow v_i\  AND\  m_{eq})$
   }{$InMemory(m_{eq}\leftarrow NOT(v_i)\  AND\  m_{eq})$
  }
 }
\caption{\small PIM controller algorithm for equality operation between an in-memory value and an immediate value.
}
\label{algo:imm_eq}
\end{algorithm}

\section{Mapping Databases to PIMDB}
\label{sec:db_with_mMPU}

In this section, we show how a database is mapped to PIMDB. First, the memory layout of a database relation is presented. Then, the PIM instructions necessary to support database applications are described. These PIM instructions need to be implemented as a sequence of bulk-bitwise logic operations by the PIM controller (described in Section~\ref{subsec:pim_module}). 

PIMDB holds a copy of subset of the database in the PIM modules to accelerate certain operations. Other operations are performed in a non-PIM manner using the host and DRAM main memory~\cite{Hasso2013}. The database copy is constructed offline once and then used for query execution. The query execution does not modify the database copy, so further data copying is not necessary prior to subsequent query execution. Note that maintaining database copies is common in database systems~\cite{Pohanka2020}.

\begin{figure}[t]
\centering
\subfloat[\label{fig:table_layout_relation}]{\includegraphics[width=4.05cm]{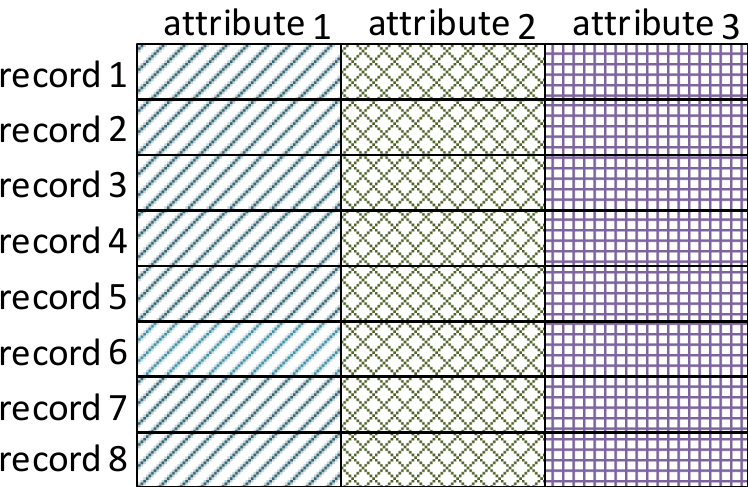}}\hspace{2em}%
\subfloat[\label{fig:table_layout_mat}]{\includegraphics[width=0.41\linewidth]{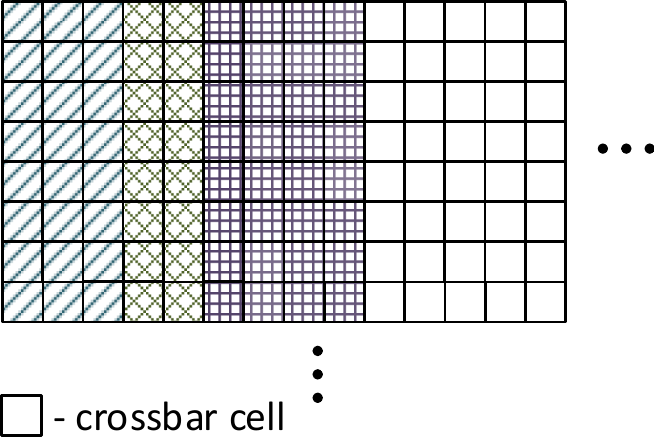}}%
\caption{(a) Visualization of a database relation. (b) The layout of the relation in a crossbar (partly shown). The attributes of a record are within a single crossbar row. Attributes are aligned across rows.}
\label{fig:table_layout}
\end{figure}

\begin{figure}[t]
\centering
\begin{minipage}{\columnwidth}
\centering
\includegraphics[trim={0cm 0cm 0cm 0},width=\columnwidth]{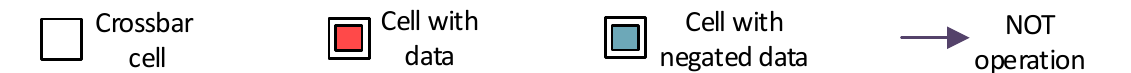}
\end{minipage}\vspace{4pt}
\subfloat[]{\includegraphics[trim={0cm 0cm 0cm 0},width=0.23\columnwidth]{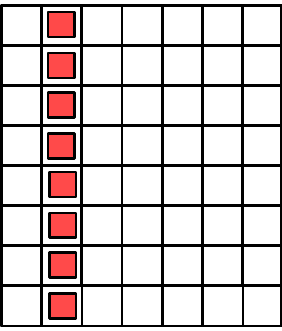}}\hspace{4pt}%
\subfloat[\label{fig:table_operation_coltrans_first_neg}]{\includegraphics[trim={0cm 0cm 0cm 0},width=0.23\columnwidth]{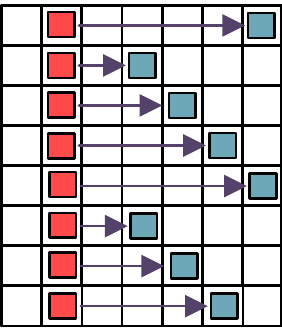}}\hspace{4pt}%
\subfloat[\label{fig:table_operation_coltrans_second_neg}]{\includegraphics[trim={0cm 0cm 0cm 0},width=0.23\columnwidth]{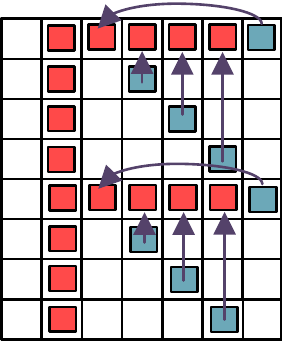}}\hspace{4pt}%
\subfloat[]{\includegraphics[trim={0cm 0cm 0cm 0},width=0.23\columnwidth]{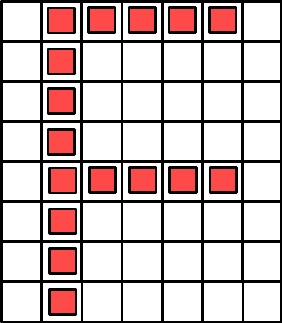}}%
\caption{Example of a column-transform operation: Transferring a single crossbar column to several rows. In this example, a column in an $8\times7$ crossbar is transformed into two rows, each with four cells. (a) The initial state of the crossbar with the input column. (b) Negating each value to its targeted result column. (c) Negating each value to its targeted result row. (d) The final state of the crossbar with input column and result.}%
\label{fig:table_operation_coltrans}%
\end{figure}

\subsection{Database Relation Memory Layout}
\label{subsec:db_layout}
\revMark{Database relations are stored in PIMDB memory such that each record is placed in a crossbar row. For each relation, enough huge-pages are assigned to it to hold all of its records.}
Records can be assigned to the rows of a crossbar in any order, and to any crossbar in the huge-pages assigned to the relation (no two relations share a page). If a relation requires more records, additional pages can be assigned to it dynamically. Each attribute is aligned in all the crossbar rows and is contained in consecutive crossbar cells\footnote{
PIM instructions are supported on consecutive cells; see Section~\ref{subsec:PIMop}.
} (Fig.~\ref{fig:table_layout} shows this mapping). Since reads and writes are performed on crossbar rows, a relation is held in a row-store manner in the PIM memory, making accessing and updating records straightforward. Nevertheless, a single record is not entirely consecutive in the host's virtual memory due to the mapping of the crossbar cells to memory addresses (see Section~\ref{subse:programmin_model}).

If relation records are too large to fit into a single crossbar row, the attributes can be distributed across several crossbars, with each crossbar on a different huge-page. 
To operate on such a relation, the host might need to transfer partial operation results between the pages through reads and writes, slowing the execution.
Note that for TPC-H, the physical size of the crossbar rows is sufficient for all relations and there is no need to split attributes of a relation among different pages.

\begin{figure}[t]
\centering
\subfloat[]{\includegraphics[trim={18pt 16pt 17pt 22pt}, clip, width=0.31\columnwidth]{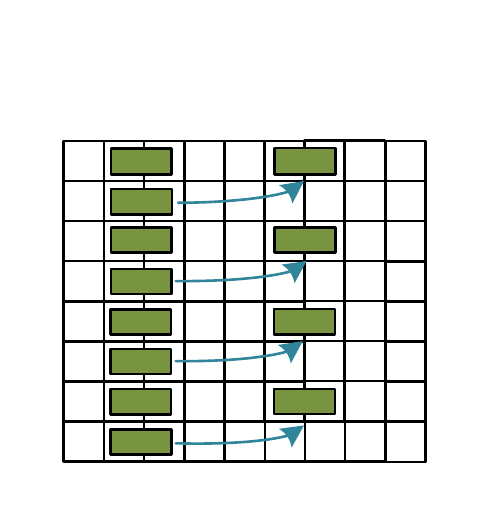}}\hspace{4pt}%
\subfloat[]{\includegraphics[trim={18pt 15pt 17pt 22pt}, clip, width=0.31\columnwidth]{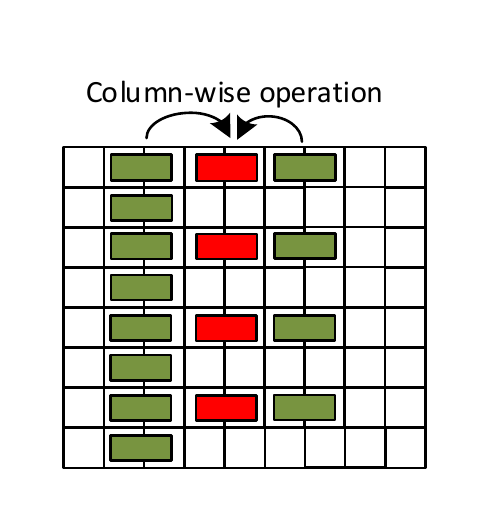}}\hspace{4pt}%
\subfloat[]{\includegraphics[trim={18pt 15pt 17pt 22pt}, clip, width=0.31\columnwidth]{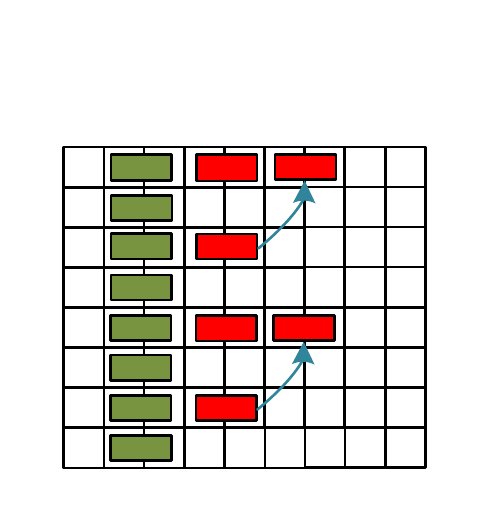}}\newline\vspace{-1pt}
\hspace{-5pt}\subfloat[]{\includegraphics[trim={18pt 15pt 17pt 20pt}, clip, width=0.31\columnwidth]{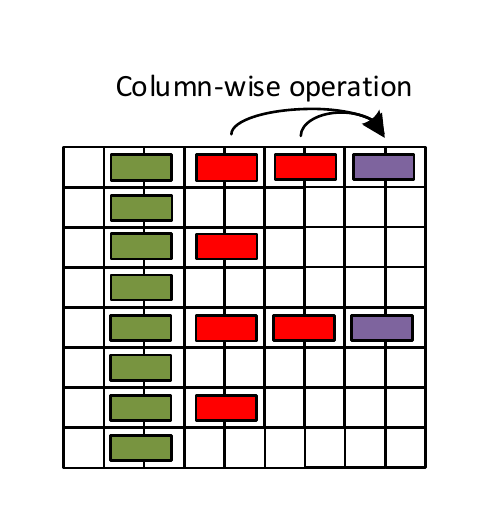}}\hspace{4pt}%
\subfloat[]{\includegraphics[trim={18pt 15pt 17pt 20pt}, clip, width=0.31\columnwidth]{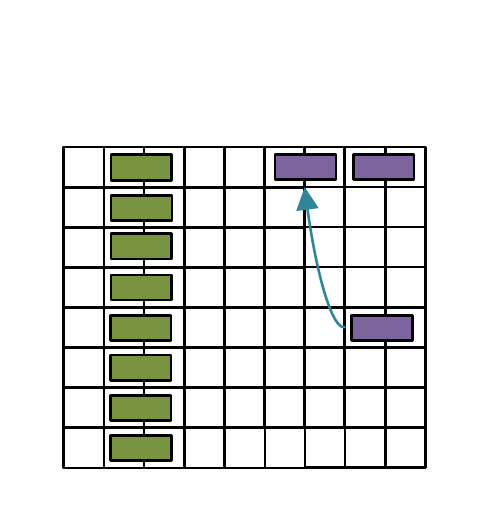}}\hspace{4pt}%
\subfloat[]{\includegraphics[trim={18pt 15pt 17pt 20pt}, clip, width=0.31\columnwidth]{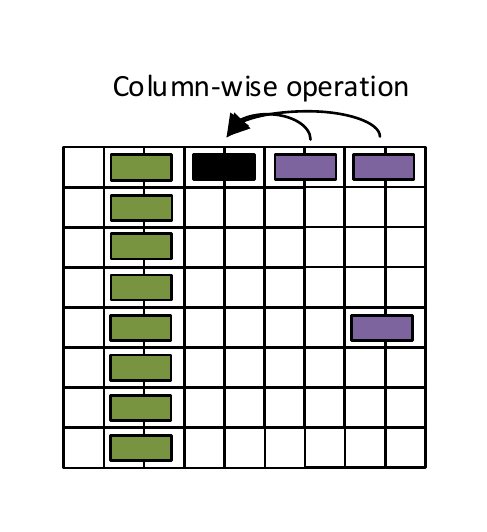}}%
\caption{Example of a reduce operation: Reducing (\myemph{SUM}/\myemph{MIN}/\myemph{MAX}) values stored in multiple rows to a single value in a single row. In this example, a column of two-cell values (green) in an $8\times8$ crossbar is reduced to a single value. Unlike the example, for some operations (\textit{e.g.}, \myemph{SUM}) the number of bits after each reduction may change. (a) Copying half of the values to the rows of the other half of the values. (b) Reducing values on the same row to a single value using a column-wise operation (\myemph{SUM}/\myemph{MIN}/\myemph{MAX}). (c) Copying half of the reduced values to the rows of the other half. (d) Reducing values again using a column-wise operation. (e) Copying half of the reduced values to the rows of the other half. (f) Reducing values for the last time, resulting in a single value.}%
\label{fig:table_operation_agg}%
\end{figure}

\subsection{PIM Instructions}
\label{subsec:PIMop}
To support database queries, PIMDB performs \textit{filter} and \textit{aggregate} operations within the crossbars. \revMark{This section explains which PIM instructions PIMDB implements to support the filter and aggregate operations. These instructions consist of traditional arithmetic instructions and three additional complex instructions, Table~{\ref{table:cycle_count}} lists all implemented instructions.}  

Filter operations operate on every record in a database relation and specify which records pass the filter condition. A filter condition is defined using the attributes of the relation and can be evaluated as \myemph{True} or \myemph{False} for each record, according to the attribute values of the record. A condition can be a single comparison or a logical combination (\emph{e.g.}, \myemph{OR, AND, NOT}) of several comparisons. A comparison is one of the following operations: $=, \ne, <, > , \leq, \geq$, between attributes, functions of attributes, and constants.

To support filter operations, the PIM controller supports comparison between two in-memory values or an in-memory value and an immediate (constant), bitwise logical operations, and simple arithmetic operations (\emph{e.g.}, addition, multiplication), all of which are performed by bulk-bitwise logic on all crossbar rows. These operations should be supported for an arbitrary length of consecutive crossbar columns since a database attribute can have any length; \emph{e.g.}, we may need to compare 8-bit numbers as well as 32-bit numbers.
When filtering using these operations, the Boolean result will appear as the value of a single cell in each crossbar row (the same cell in all rows, indicated by the PIM request address and constituting a crossbar column). The Boolean result indicates whether or not the record associated with that row passed the filter condition. Thus, by performing filter operations in memory, instead of reading all the attributes for filtering per record, only a single bit per record is read, substantially reducing the data transfer.

Since the reads from the crossbar are row-oriented, the PIM controller also supports a \emph{column-transform} instruction to enable efficient retrieval of the filter result. The column-transform operation takes a single cell column at each crossbar, and transfers the values of the column's cells into several crossbar rows. The number of values in each target row is fixed to the size of a read from the crossbar. After transforming the filter results into a row orientation, the results can be read with higher efficiency than with a column orientation. The column-transform operation is shown in Fig.~\ref{fig:table_operation_coltrans}.

\begin{table}[t]
\setlength\tabcolsep{3pt}
\centering
\begin{tabular}{|c|c|c|c|c|c|} 
\hline
\rule{0pt}{1.5ex} Relation	          & In & \# of      & \# of Crossbar & \# of PIM & Memory \\
\rule{0pt}{1.5ex}  	          &PIM&  Records    &  Row Bits & Pages   & Utilization \\\hline
\rule{0pt}{1.5ex} \myemph{PART}     & \checkmark    & $2\times10^8$ 	 & 124                & 12                      & 24.1\%\\ \hline
\rule{0pt}{1.5ex} \myemph{SUPPLIER} & \checkmark    & $1\times10^7$ 	 & 99                 & 1                       & 12\%\\\hline
\rule{0pt}{1.5ex} \myemph{PARTSUPP} & \checkmark	   & $8\times10^8$	 & 80                 & 48                     & 15.5\%\\\hline
\rule{0pt}{1.5ex} \myemph{CUSTOMER} & \checkmark	   & $1.5\times10^8$ & 106                & 9                       & 20.6\%\\\hline
\rule{0pt}{1.5ex} \myemph{ORDERS}   & \checkmark    & $1.5\times10^9$ & 133                & 90                   & 25.8\%\\ \hline
\rule{0pt}{1.5ex} \myemph{LINEITEM} & \checkmark    & $6\times10^9$	 & 191                & 358                & 37.3\%\\ \hline
\rule{0pt}{1.5ex} \myemph{NATION}   & X     & 25	             & -                  & -                      & -\\ \hline
\rule{0pt}{1.5ex} \myemph{REGION}  & X     & 5	             & -                  & -                      & -\\ \hline
\rule{0pt}{1.6ex} \textbf{\myemph{Total}}  & -     & -	             & -                  &  \textbf{518}                      &  \textbf{32.6\%}\\ \hline
\end{tabular}
\vspace{3pt}
\caption{PIM layout summary for TPC-H relations with $SF=1000$}
\label{table:TPCH}
\end{table}
\setlength\tabcolsep{6pt}

Aggregate operations are a reduction of many values to a single value, \emph{e.g.}, taking the sum/average/minimum of an attribute from selected records in a relation. To support aggregation, the PIM controller implements \textit{reduce} instructions on consecutive crossbar columns across all rows, as shown in Fig.~\ref{fig:table_operation_agg}. The cells in each crossbar row for the given consecutive columns are assumed to constitute a single value, and the values from all rows are reduced to a single value result according to the specific arithmetic operation. The result location is specified by the PIM request. If a value of a specific row needs to be ignored, it should be adjusted beforehand by PIM operations according to the specific reduce operation. For instance, to ignore
rows in a \myemph{SUM} reduce, a filter should be computed and \myemph{AND} with the column value to be reduced. This will zero undesirable values 
and the \myemph{SUM} will operate on the desired values only.
After the reduce operation, the reduced values from all crossbars are read and combined by the host processor. Thus, when performing filtering and aggregation, instead of reading the attributes for each record, only a single value is read from each crossbar per aggregation (\textit{e.g.}, in our evaluation, 1024 records per crossbar are reduced to a single value, \revMark{different number of records in a crossbar changes the read reduction amount}).

Since reduce operates first on values in the same crossbar, and then on values across crossbars, only commutative and associative operations can be supported, such as \myemph{SUM}, \myemph{MIN}, and \myemph{MAX}. Non-commutative or non-associative aggregate operations (\emph{e.g.}, average) can be implemented using several supported instructions and additional host computation. For example, to perform an average, the PIM module performs a \myemph{SUM} on the requested attribute and then another \myemph{SUM} on the filter result (in a column orientation, providing the record count). Thereafter, the host performs the division between the two \myemph{SUM} results.

As stated before, PIM instructions operate on crossbar columns, and relation attributes are set along crossbar columns. Thus from the PIM perspective, the relation is held in PIMDB as a column-store. Hence, reads and writes see relations in a row-store (Section~\ref{subsec:db_layout}), while PIM operations see a column-store. 

\section{Evaluation Methodology}
To evaluate PIMDB, the gem5 full-system simulator~\cite{gem5} is used with the necessary additions and modifications. As benchmarks, TPC-H~\cite{TPCH} queries are used.
In this section, we elaborate on PIMDB and workload implementation details, including the database layout in the memory, query execution, and micro-architecture details. Following this, the evaluation results are presented and discussed.

\subsection{TPC-H Benchmark}
\label{subsec:TPCH}
TPC-H~\cite{TPCH} is a database benchmark for business decision support, which specifies a database and a suite of queries.
TPC-H controls the size of the database through a \emph{scale factor (SF)}. The size of some relations is linearly dependent on the $SF$, and the database total size is approximately the $SF$ in units of GB. 
We evaluate the proposed design with $SF=1000$, \textit{i.e.}, a database approximately 1TB in size (before compression). 

\begin{table}[t]
\setlength{\tabcolsep}{2pt}
\centering
\begin{tabular}{|c|D||c|R|} 
\hline  	          \multicolumn{4}{|c|}{\rule{0pt}{1.6ex}Filter-Only Queries} \\\hline
\rule{0pt}{1.5ex} \myemph{Q2}  & \myemph{PART,SUPPLIER}                         &  \myemph{Q3}   & \myemph{CUSTOMER, ORDERS,LINEITEM}  \\ \hline
\rule{0pt}{1.5ex} \myemph{Q4}  & \myemph{ORDERS, LINEITEM}             &  \myemph{Q5}   & \myemph{SUPPLIER, CUSTOMER,ORDERS} \\ \hline
\rule{0pt}{1.5ex} \myemph{Q7}  & \myemph{SUPPLIER, CUSTOMER,LINEITEM} & \myemph{Q8}    & \myemph{PART,ORDER, CUSTOMER} \\\hline
\rule{0pt}{1.5ex} \myemph{Q10} & \myemph{ORDERS, LINEITEM}             &  \myemph{Q11}  & \myemph{SUPPLIER}\\\hline
\rule{0pt}{1.5ex} \myemph{Q12} & \myemph{LINEITEM} 	                 & \myemph{Q14}   & \myemph{LINEITEM} \\\hline
\rule{0pt}{1.5ex} \myemph{Q15} & \myemph{LINEITEM}                     & \myemph{Q16}   & \myemph{PART} \\\hline
\rule{0pt}{1.5ex} \myemph{Q17} & \myemph{PART}	                     & \myemph{Q19}	  & \myemph{PART,LINEITEM} \\\hline
\rule{0pt}{1.5ex} \myemph{Q20} & \myemph{SUPPLIER, LINEITEM}	         & \myemph{Q21}   & \myemph{SUPPLIER, ORDERS,LINEITEM} \\\hline
\multicolumn{4}{|c|}{\rule{0pt}{1.6ex}Full Queries} \\\hline
\rule{0pt}{1.5ex} \myemph{Q1}   & \myemph{LINEITEM}    & \myemph{Q6}   & \myemph{LINEITEM}  \\ \hline
\rule{0pt}{1.5ex} \myemph{Q22\_sub} & \myemph{CUSTOMER}    & -	 & -  \\ \hline
\end{tabular}
\vspace{3pt}
\caption{PIM operated relations for each query in our evaluation. Filter-only queries may operate on more relations in a non-PIM manner.}
\label{table:queries}
\end{table}

Table~\ref{table:TPCH} summarizes the layout of the TPC-H relations in the PIM modules.
Relations with few records are not assigned to the PIM modules. Instead, they are column-stored in DRAM, as directly accessing a few records in DRAM is more efficient than PIM operations.
For PIM module relations, attributes are compressed using simple schemes, without limiting the relevant PIM operations, when applicable: dictionary encoding~\cite{Hasso2013} (allowing equality comparisons), or leading-zero suppression~\cite{Daniel06} (allowing all operations).
Large text attributes that are almost unused by the TPC-H queries are not included in the PIM module database copy; this is done to preserve space for computation. These attributes are the \myemph{NAME}, \myemph{ADDRESS}, and \myemph{COMMENT} of the various relations. If a query requires the use of these attributes, they are accessed from the DRAM. Additionally, a \myemph{valid} attribute is added to each relation in the PIM module so that unused crossbar rows can be ignored in query execution.
Decisions on which relations should reside in the PIM module, what attributes to include, and how to encode them, were made manually for the evaluation presented here. Such decisions can be automated in the same manner as other implementation decisions in database management~\cite{Dageville2004}.

The memory utilization listed in Table~\ref{table:TPCH} considers the data size of the relations in the PIM module and the allocated memory space in huge-pages, capturing the memory overhead for enabling PIM. The relatively low memory utilization is primarily due to the low utilization of a crossbar row, rather than unused rows, \textit{i.e.}, most of the unoccupied space can be used for intermediate results during the computation and is not wasted. Furthermore, this space can be used for extending the life of memory cells with wear leveling (see Section~\ref{subsec:endurance}).

Using the TPC-H benchmark, we evaluate the query operations that can be performed with the PIM modules, as the focus is on bulk-bitwise operation performance.  
Three of the queries, \myemph{Q1}, \myemph{Q6}, and a sub-query of \myemph{Q22} (\myemph{Q22\_sub}), operate on single relations only, so both filtering and aggregation can be performed in the PIM modules. The rest of the queries operate on multiple relations and can perform only the filter part in the PIM module. These two groups are referred to as \emph{full queries} and \emph{filter-only queries}, respectively. Three of the filter-only queries, \myemph{Q9}, \myemph{Q13}, and \myemph{Q18}, only filter attributes that are not included in the PIM memory (see the start of this section), so they do not involve any bulk-bitwise operations and are not evaluated. Table~\ref{table:queries} presents a by-query summary of the relations that are operated on using the PIM modules (Table~\ref{table:queries} only specifies relations involving PIM operations).

\subsection{PIM Module Micro-Architecture}
\label{subsec:PIM_Module_Evaluated_Details}
\begin{table}[t]
\setlength{\tabcolsep}{2pt}
\centering
\begin{tabular}{|A|B||E|F|} 
\hline
\multicolumn{4}{|c|}{\rule{0pt}{1.6ex}Single PIM Module (8 PIM Modules)}\\
\hline
\rule{0pt}{1.5ex}   Capacity & \mbox{128GB (1TB)}  & Banks  &64 (512)  \\\hline
\rule{0pt}{1.5ex}   Subarrays per \mbox{PIM controller}    & 64     & Crossbars per subarray  & 4 \\ \hline
\rule{0pt}{1.5ex}   Crossbar rows  & 1024  & Crossbar columns  & 512   \\\hline
\rule{0pt}{1.5ex}   Crossbar read  & 16 bit  & Stateful logic cycle  &  30 ns~\cite{CONCEPT}  \\\hline
\rule{0pt}{1.5ex}   Crossbar write energy  & \mbox{6.9 pJ/bit} \cite{CONCEPT}  & Single stateful logic energy  &  \mbox{81.6 fJ/bit} \cite{Talati2016} \\\hline
\rule{0pt}{1.5ex}   Crossbar read energy  & \mbox{0.84 pJ/bit} \cite{CONCEPT}  & Single PIM controller power  &  126 uW \\\hline

\multicolumn{4}{|c|}{\rule{0pt}{1.6ex}Evaluation System}\\
\hline 
\rule{0pt}{1.5ex}Processor Cores& \mbox{\revMark{6} cores, X86,} \mbox{OoO, \revMark{3.6GHz}}  & Main memory  & \mbox{\revMark{64GB} DRAM}, \mbox{DDR4-2400}, \mbox{\revMark{2 channels}}  \\\hline
\rule{0pt}{1.5ex}   L1 cache & \mbox{Private, \revMark{64KB},} \mbox{64B block}, 4-way  & L2 cache & \mbox{Shared, \revMark{8MB},} \mbox{64B block}, 16-way \\\hline
\rule{0pt}{1.5ex} Coherence protocol  & MESI  & PIM modules & 8 \\\hline

\end{tabular}
\vspace{3pt}
\caption{Architecture and system configuration}
\label{table:arch_details}
\end{table}

A single PIM module is a single memory rank, comprising a media controller chip and eight PIM-enabled memory chips. The capacity of a single memory rank is 128GB (equivalent to the capacity of the small Intel Optane module~\cite{OptaneDC}). The details of the PIM module are listed in Table~\ref{table:arch_details} and explained here.

\subsubsection{Interfaces}
The operation timing for the R-DDR interface, between the media controller and the memory chips, is taken from~\cite{CONCEPT}.
The media controller communicates with the host memory controller using the OpenCAPI protocol~\cite{OpenCAPI}. Our simulation implementation of the OpenCAPI protocol considers the added protocol header sizes and computes the timing according to the OpenCAPI bandwidth of 25GB/s~\cite{OpenCAPI}.

\subsubsection{Instruction Implementation}
The PIM controller supports the operations listed in Table~\ref{table:cycle_count}, performed on all crossbars connected to the controller in parallel.
The \textit{column-transform}, shown in Fig.~\ref{fig:table_operation_coltrans}, instruction is implemented by negating the source column into each destination column (Fig.~\ref{fig:table_operation_coltrans_first_neg}), and negating each bit a second time within its designated column to its designated row (Fig.~\ref{fig:table_operation_coltrans_second_neg}). The \textit{reduce} instructions, shown in Fig.~\ref{fig:table_operation_agg}, are implemented as iterations of move and reduce steps, resulting in a binary-tree reduction scheme. In a move step, half of the values are moved to the rows of the other half, while in the reduce step, the two values in each row are reduced according to the relevant operation. Hence, each iteration reduces the number of values by half. The process stops when a single value is left. All other instructions are implemented as iterations on a single-bit  operation (Section~\ref{subsec:InstructionImpelemntation}).

\subsubsection{Crossbar Operations}
\label{subsubsec:crossbar_operations}
We take a conservative and practical approach regarding the capabilities of bulk-bitwise logic operations in a crossbar, \textit{i.e.}, the operations sent by a PIM controller to a crossbar. We impose restrictions on bulk-bitwise logic operations to minimize the area overhead of the crossbar peripherals and communication from the PIM controllers to the crossbars. This overhead is due to the additional voltage levels added for each enabled stateful logic operation~\cite{MAGIC}, which must be multiplexed on bitlines and wordlines, and for adding crossbar peripherals. In addition, allowing full flexibility to the combination of operated crossbars rows and columns in a bulk-bitwise logic operation, as in~\cite{SIMPLER}, requires controlling each row or column independently. This requires passing a wire for each wordline and bitline (1536 wires in our design) from a PIM controller to its crossbars. As there are many crossbars per PIM controller, the required wiring can reduce the memory density and increase area. 

To address this challenge, we enable only a minimal number of operation types and a minimum allowed combination of crossbar rows and columns, while still maintaining complete functionality. First, column-wise operations can be \myemph{NOR2}, \myemph{NOT}, single-column-\myemph{SET}, or single-column-\myemph{RESET} and can only be performed on all the rows in parallel. Exclusion of rows from computing should be done through software, \emph{e.g.}, using a bit-mask and including it in the computation. Second, row-wise operations can be performed only on a single column at a time and can either be a \myemph{NOT} or single-row-\myemph{SET}~\cite{Talati2016}. These minimum capabilities are essential for a functionally complete column-wise computation, maintaining high parallelism, and enabling data movement between rows (for column-transform and reduce instructions). These restrictions make certain optimizations impossible, posing an interesting topic for future research. For example, in reduce instructions, the reduce steps are performed on all rows, including rows not participating in a step. The resulting instruction cycle count and intermediate cell requirements (computed as in~\cite{Talati2016}) are listed in Table~\ref{table:cycle_count}.

\begin{table}[t]
\centering
\begin{tabular}{|c|c|c|} 
\hline
\rule{0pt}{1.4ex}\bf{Instruction}&\bf{Cycle Count}&\bf{Inter. Cells}\\\hline
\rule{0pt}{1.4ex}Equal imm          & $imm_0+3\cdot imm_1+1$    &$1$\\\hline
\rule{0pt}{1.4ex}Not Equal imm      & $imm_0+3\cdot imm_1+3$            &$2$\\\hline
\rule{0pt}{1.4ex}Less Than imm      & $11\cdot imm_0+3\cdot imm_1+4$    &$5$\\\hline
\rule{0pt}{1.4ex}Greater Than imm   & $11\cdot imm_0+3\cdot imm_1+2$    &$6$\\\hline
\rule{0pt}{1.4ex}Add imm            & $18n+3$   & $8$\\\hline
\rule{0pt}{1.4ex}Equal       & $11n+3$   & $5$\\\hline
\rule{0pt}{1.4ex}Less Than   & $16n+2$   & $6$\\\hline
\rule{0pt}{1.4ex}Set/Reset          & $n$       & $0$\\\hline
\rule{0pt}{1.4ex}Bitwise NOT        & $2n$      & $0$\\\hline
\rule{0pt}{1.4ex}Bitwise AND        & $6n$      & $2$\\\hline
\rule{0pt}{1.4ex}Bitwise OR         & $4n$      & $1$\\\hline
\rule{0pt}{1.4ex}Addition           & $18n+1$   & $6$\\\hline
\rule{0pt}{1.4ex}Multiply           & $24nm-19n+2m-1$        & $6$\\\hline
\rule{0pt}{1.4ex}\textbf{Reduce Sum}         & $2254n+3006$ & $n+15$\\\hline
\rule{0pt}{1.4ex}\textbf{Reduce Min/Max}     & $2306n+200$ & $n+7$\\\hline
\rule{0pt}{1.4ex}\textbf{Column-Transform}    & $2050$ & $1$\\\hline
\end{tabular}
\vspace{3pt}
\caption{Instruction characteristics. $n$, $m$: operands and immediate lengths. $imm_0$/$imm_1$: immediates' number of 0/1 bits. Intermediate cells: cells required for intermediate results in addition to input and output cells per crossbar row. Crossbar size affects bold-marked operations' coefficients; values in the table reflect a crossbar size of $1024\times 512$.
}
\label{table:cycle_count}
\end{table}

\subsection{System Configuration}
\label{subset:system}
As the reduction of memory reads is done by the PIM modules and the evaluated workload is memory constrained, the choice of a host (\textit{e.g.}, CPU, GPU, FPGA, \revMark{near-memory processing}) will primarily affect the memory accessing speed. Thus, although the choice of the host will affect performance, it will not change the number of memory reads that are eliminated due to the PIM execution. Hence, for simplicity of evaluation, we chose a multicore with \revMark{six} out-of-order X86 cores as the host, having private L1 caches and a shared L2 cache. The host has a \revMark{64GB} DDR4-2400 DRAM main memory, along with eight OpenCAPI channels\footnote{IBM's Power9 has 16 OpenCAPI channels~\cite{IBMPower9}.}, each with a single PIM module, totaling 1TB\revMark{\footnote{\revMark{A single PIM module has a capacity of 128GB, the same capacity as the smallest Intel Optane module}{\cite{OptaneDC}}.}} of memory in the PIM modules. Table~\ref{table:arch_details} summarizes the system configuration.

The system was implemented in the gem5 simulation~\cite{gem5}, running in full-system mode (running a Linux kernel), and includes the gem5's ruby cache system. 
A PIM request instruction was added to the host instruction set.
Similarly to a write, a PIM request instruction requires two registers as input: address and data.

\subsection{Query Compilation and Execution}
\label{subsec:compile}
We built an SQL compiler to abstract PIMDB and its programming model, allowing it to be programmed using a high-level language. The compiler receives the configuration of the database and the SQL statements as input and generates a C++ code for the query execution, where PIM requests are issued using an inline assembly. The C++ code is then compiled using gcc. This compilation process is done offline and is not measured as part of the query execution.

The compiler extracts the query operations required for the PIM modules' execution, \emph{i.e.}, filtering the relations, aggregating if possible, and reading the results. The query execution starts by operating on the small relations residing in the DRAM memory, 
if required. 
For relations in the PIM module, the operation execution is divided among four threads, one per core.
Each receives part of the huge-pages for each relation; thus, each thread operates on separate pages. For each relation participating in the query, each thread performs several computations and read phases (possibly one of each). A computation phase sends the necessary PIM requests to the thread's pages and the read phase reads the results from those pages. Each computation phase fits into the available crossbar area unoccupied by data, and following it, a read phase clears the area for reuse. To avoid unnecessary cache flushes, the compiler assigns the results of different computation phases to different cache blocks by controlling the addresses of the PIM requests. Memory ordering among PIM requests and reads is enforced using memory fences. Phases of two relations are not interleaved in each thread.  

Before query execution is simulated, the required huge-pages are first allocated using the operating system and then populated according to~\cite{TPCH}. To maintain reasonable memory requirements for the simulation resources, the 1GB huge-pages are emulated using 2MB pages.
For the emulation, the number of allocated 2MB pages is the same as the required number of 1GB pages. To emulate the execution time, the required number of reads from each 2MB page in the query execution is matched to the required number of reads on a 1GB page (we made sure each read is retrieved from the PIM modules and not from the caches). Since the latency of PIM operations is independent of the page size, it does not need adjustments. For energy and power considerations, however, the relevant operation energies were counted as operating on 1GB pages, (\emph{e.g.}, number of bulk-bitwise logic operations, number of PIM controllers, \emph{etc.}).

\subsection{Baseline}
To show the bulk-bitwise PIM benefits, we take as a baseline the execution of the same query operations evaluated on PIMDB, performed on a database contained in the host main memory (\textit{i.e.}, \textit{in-memory database}~\cite{Hasso2013}). The baseline is executed on the same PIMDB host system configuration, using the gem5 full-system simulation. Comparing our system to the same host system without PIMDB demonstrates the memory access reduction achieved with bulk-bitwise PIM operations. For the baseline, all database relations are column-stored and use the same encoding and compression techniques as in PIMDB. For each query, a C++ code was written to perform the same query operations as produced by our PIMDB compiler. The relevant attributes for each query are brought into the main memory prior to execution (no disk access during execution). The same huge-page emulation is used for memory allocation as in the PIMDB evaluation. Hence, the baseline represents operations on a column-store in-memory database~\cite{Hasso2013} equivalent to the operations executed in PIMDB. \revMark{The execution of these operations, however, cannot be easily broken to its different components (host compute, PIM compute, memory reads) as the PIMDB execution since the host executes out-of-order and the operation of these components overlap in time.}

The baseline execution structure is also similar to that of PIMDB. The small tables (\emph{i.e.}, \myemph{REGION} and \myemph{NATION}) are operated on if necessary, and then the large relations operation is split into four threads where each thread is responsible for a quarter of each relation's records. The threads serially traverse the records of the required relations, filtering according to the required attributes (using nested if-statements) and performing aggregation. The attribute filtering order is chosen offline to minimize memory access to non-required records. 

\section{Results}
\label{sec:results}

\subsection{Query Execution Time}
\label{subsec:query_execution}
To evaluate the benefits of PIMBD, we measure the query execution time, capturing the operations that PIMDB accelerates, and compare it to the query execution time of the baseline.
PIMDB execution time includes the operations performed by the PIM modules, the operations to support the PIM operations (such as operations on small DRAM relations, spawning threads, \emph{etc.}), and the results read from the PIM module. Hence, for the full queries, the entire query execution is captured. For the filter-only queries, however, only the filter operation is measured (including reading the results), as the rest of the query execution does not involve the PIM modules or their database copy and is considered out-of-scope for this work.
Our evaluation does not compare PIMDB to other bulk-bitwise PIM technologies and architectures~\cite{AMBIT,CONCEPT,Pinatubo,SIMDRAM,Giannopoulos2020,NVQuery,RACER} as we are interested in evaluating bulk-bitwise PIM, not comparing the various technologies. We discuss this issue further in Section~\ref{sec:related_work}. 

\begin{figure}[t]
\centering%
\subfloat[\label{fig:speedup_filter}Filter-only queries]{\includegraphics[width=0.66\linewidth]{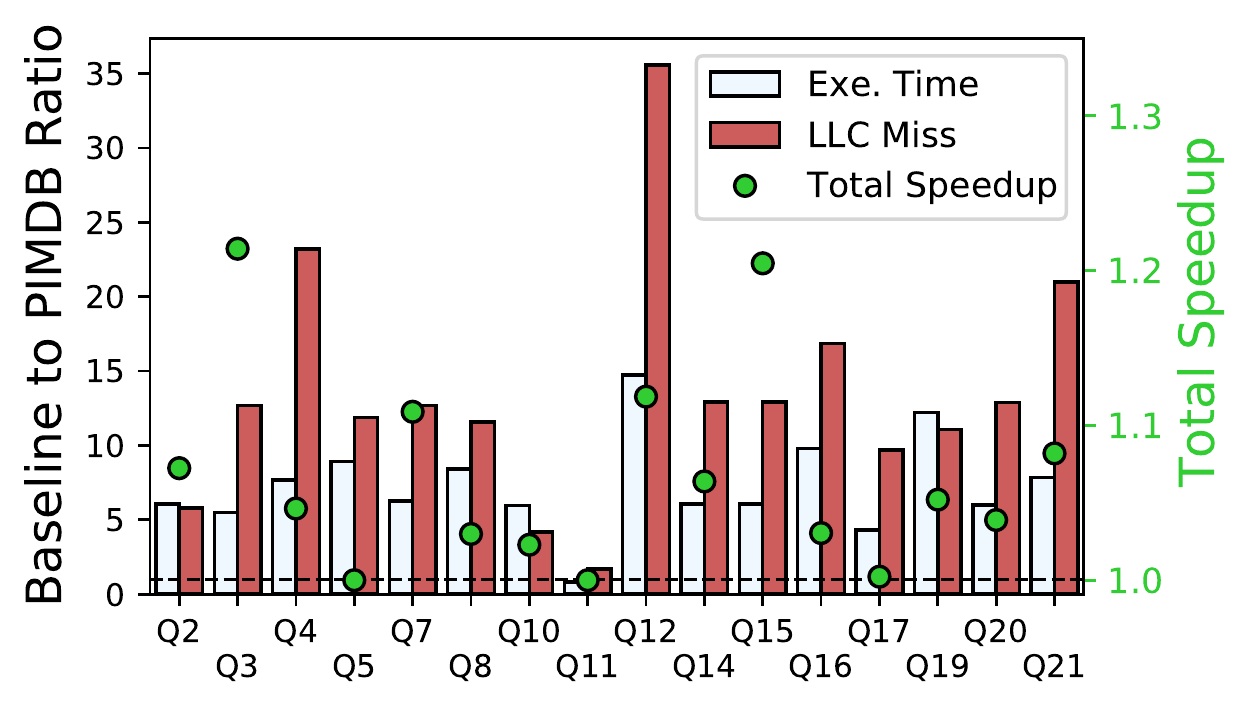}}\hspace{3pt}%
\subfloat[\label{fig:speedup_full}Full queries\\\revMark{(filter+aggregation)}]{\includegraphics[width=0.275\linewidth]{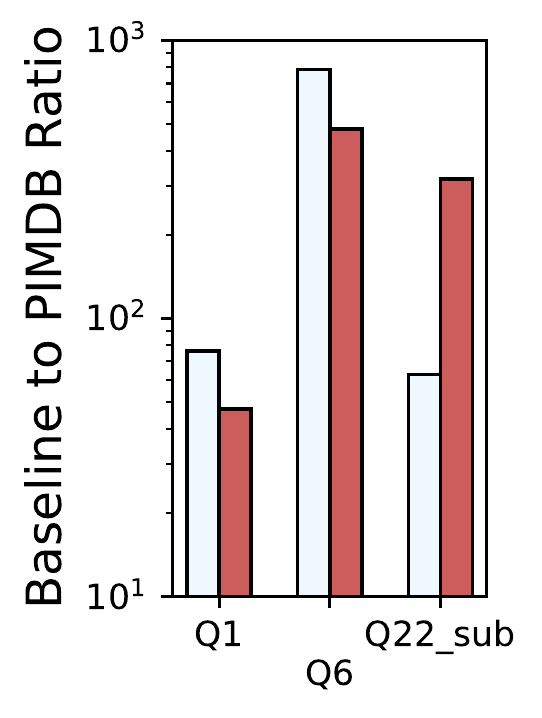}}%
\caption{PIMDB execution time speedup and reduction in LLC misses compared to the baseline (left y-axis of sub-figures). An estimated total query speedup for the filter-only queries is shown using the right y-axis of sub-figure (a).} 
\label{fig:speedup}%
\end{figure}

Fig.~\ref{fig:speedup} shows the execution time and the LLC miss ratios between PIMDB and the baseline on the accelerated operations. The filter-only and full queries achieve a speedup of \revMark{$0.82\times$--$14.7\times$ and $62\times$--$787\times$}, respectively. 
The speedup difference between full and filter-only queries is due to the data reduction advantage of reduce versus filter operations (Section~\ref{subsec:PIMop}). \myemph{\revMark{Q11}}\revMark{ is the only query where using PIMDB results in a slowdown. }\myemph{\revMark{Q11}}\revMark{ has only a small filter operation on a small relation, achieving little read reduction and making the read latency of the results (from DRAM in the baseline and from PIM memory in PIMDB) more prominent.}

The reduction in memory accesses is perceived through the LLC misses. The LLC miss and execution time ratios, however, do not correlate entirely. The LLC misses are for all the memory accesses (not just the database) and code execution and memory access patterns for PIMDB and baseline differ substantially. Given that PIMDB accelerates only the filter operations on the filter-only queries, to provide a full performance perspective, Fig.~\ref{fig:speedup}(a) shows the total query speedup for the filter-only queries using data from~\cite{Kepe2019}. As our work aims to enable and investigate the capabilities of bulk-bitwise PIM, we focus on the parts accelerated by PIMDB and leave advanced algorithm implementation (\textit{e.g.}, JOIN, UPDATE, GROUPBY~\cite{Hasso2013}) and mapping of filter-heavy databases (\emph{e.g.}, key-value store~\cite{Pilman2017}) for future work.

\begin{figure}[t]%
\centering
\subfloat[Filter-only queries]{\includegraphics[width=\fullwidt\linewidth]{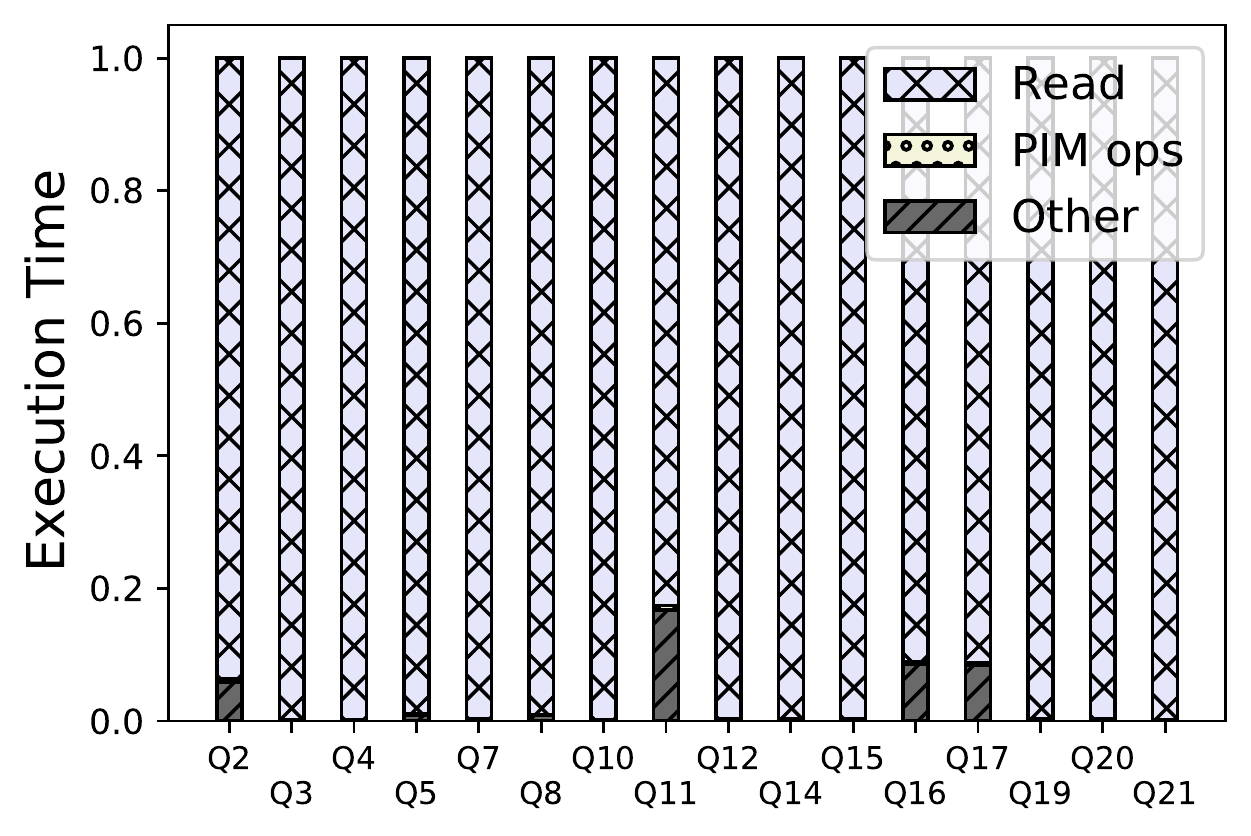}}\hspace{23pt}%
\subfloat[Full queries\\\revMark{(filter+aggregation)}]{\includegraphics[width=\filterwidt\linewidth]{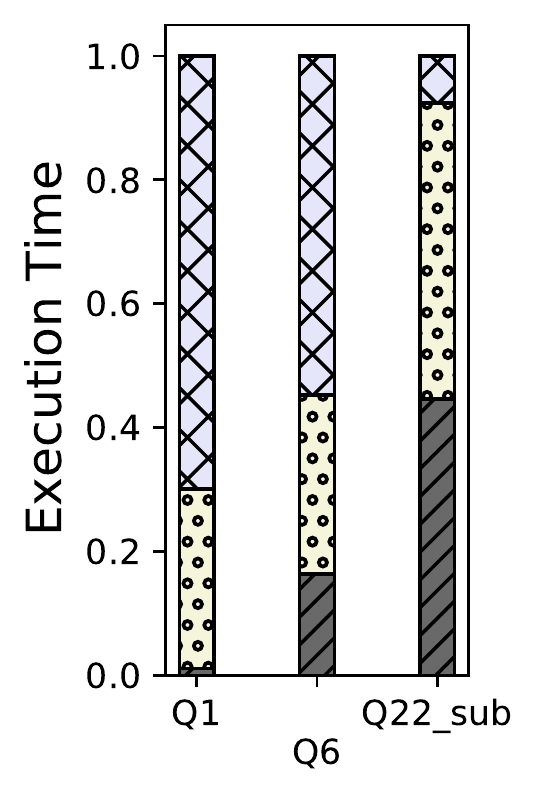}}%
\caption{PIMDB execution time breakdown. For the filter-only queries, the PIM ops portion is too small to be visible.} %
\label{fig:runtime_breakdown}%
\end{figure}

\newcolumntype{K}{>{\centering\arraybackslash}m{19 pt}}
\begin{table}[t]
\centering
\setlength{\tabcolsep}{1.2pt}
\begin{tabular}{|K|J|J|H|G||c|J|J|H|G|} 
\hline
\rule{0pt}{1.5ex}&\bf{Filter}&\bf{Arith.}&\bf{Col. trans.}&\bf{Inter. cells}&&\bf{Filter}&\bf{Arith.}&\bf{Col. trans.}&\bf{Inter. cells}\\\hline
\rule{0pt}{1.5ex}\myemph{Q2}&619&0      &2050&80&\myemph{Q3}  &97&0   &2050&32\\\hline
\rule{0pt}{1.5ex}\myemph{Q4}&216&0      &2050&49&\myemph{Q5}  &220&0  &2050&33\\\hline
\rule{0pt}{1.5ex}\myemph{Q7}&200&0      &2050&30&\myemph{Q8}  &200&0  &2050&31\\\hline
\rule{0pt}{1.5ex}\myemph{Q10}&220&0     &2050&33&\myemph{Q11} &22&0   &2050&30\\\hline
\rule{0pt}{1.5ex}\myemph{Q12}&678&0     &2050&39&\myemph{Q14} &252&0  &2050&39\\\hline
\rule{0pt}{1.5ex}\myemph{Q15}&228&0     &2050&39&\myemph{Q16} &271&0  &2050&48\\\hline
\rule{0pt}{1.5ex}\myemph{Q17}&37&0      &2050&32&\myemph{Q19} &606&0  &2050&64\\\hline
\rule{0pt}{1.5ex}\myemph{Q20}&220&0     &2050&39&\myemph{Q21} &216&0  &2050&30\\\hline\hline
\rule{0pt}{1.5ex}&\bf{Filter}&\bf{Arith.}&\bf{Agg. col/row}&\bf{Inter. cells}&&\bf{Filter}&\bf{Arith.}&\bf{Agg. col/row}&\bf{Inter. cells}\\\hline
\rule{0pt}{1.5ex}\myemph{Q1}&190&20498&$2.2\times10^{5}$/ $2\times10^{6}$&313&\myemph{Q6}&346&3390&$9.9\times10^{3}$/ $9.4\times10^{4}$&189\\\hline
\rule{0pt}{1.5ex}\myemph{Q22 \_sub}&453&106&$6.2\times10^{3}$/ $4.9\times10^{4}$&122&&&&&\\
\hline
\end{tabular}
\vspace{3pt}
\caption{The number of PIM bulk-bitwise logic cycles by type and intermediate results' cells used on a single crossbar. Filter-only queries do not use aggregation; full queries do not use column-transform. The Agg. col/row column shows the cycles of the aggregation column and row operations, respectively.}
\label{table:cycle_breakdown}
\end{table}

For calibration, we estimate how bulk-bitwise PIM compares with real systems. We compared PIMDB with the two top-ranking state-of-the-art systems for TPC-H with $SF=1000$~\cite{TPCH,Dell-TPCH,Dell-TPCH2}. These systems have a higher core count than our base system. The published results, however, are for full queries only, allowing us to compare only \myemph{Q1} and \myemph{Q6}. 
On \myemph{Q1}, we attain a speedup of \revMark{$9.3\times$ and $8.2\times$}, while on \myemph{Q6}, we achieve a speedup of \revMark{$19.6\times$ and $11.6\times$}, over~\cite{Dell-TPCH} and~\cite{Dell-TPCH2}, respectively. Note that this comparison is between simulation and real hardware, giving only a rough estimation of the PIMDB system benefits. Also we point out that taking either~\cite{Dell-TPCH} or~\cite{Dell-TPCH2} as the PIMDB host system may further improve the performance of PIMDB, as it will likely benefit from the higher core count of these systems.

Fig.~\ref{fig:runtime_breakdown} shows the breakdown of the PIM-executed queries into PIM operations, reading data from the PIM modules, and other operations (\emph{e.g.}, spawning threads, operating on DRAM relations). The breakdown shows that for the filter-only queries, the read time from the PIM modules dominates, and comprises over 99\% of execution time for all queries, with the exception of \myemph{Q2}, \myemph{Q11}, \myemph{Q16}, and \myemph{Q17} due to operation on smaller relations (\emph{i.e.}, not operating on the LINEITEM or ORDERS relations). For the full queries \myemph{\revMark{Q1}}\revMark{ and }\myemph{\revMark{Q6}}, the read time is also the bottleneck, but more moderately, \revMark{70\%}\revMark{ and }\revMark{55\%} of execution time\revMark{, respectively}. \revMark{For the full query }\myemph{\revMark{Q22\_sub}}\revMark{, due to the small read size resulting from the aggregation read reduction and the operation on the smaller }\myemph{\revMark{CUSTOMER}}\revMark{ relation, the read time is no longer the bottleneck.} These results imply that although the data transfer bottleneck was successfully relaxed, \revMark{for most cases,} it still remains the main component of the query execution time.

\begin{figure}[t]
\centering
\includegraphics[width=0.80\linewidth]{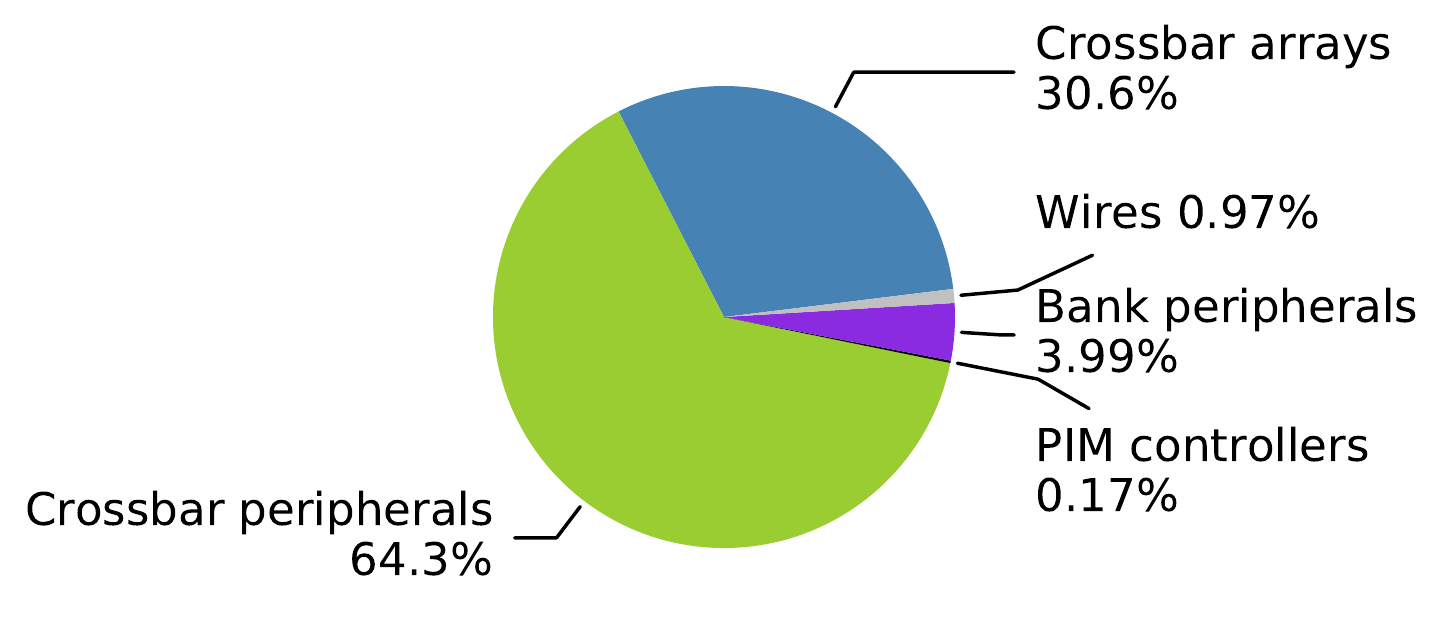}%
\caption{PIM module chip area breakdown \revMark{was taken using NVSim~}\cite{NVSim}\revMark{. Crossbar peripherals include row decoders, column multiplexers, sense amplifiers, and write drivers~}\cite{NVSim}.} 
\label{fig:area}
\end{figure}

To further investigate the PIM operations, Table~\ref{table:cycle_breakdown} shows the breakdown of bulk-bitwise PIM logic cycles by type and the number of cells used to hold intermediate results. 
The dominant operations are the column-transform operations in the filter-only queries and the aggregate operations in the full queries. \revMark{Although these operations may consume many cycles, they are amortized over all the records in many huge pages operating concurrently, where each such page ($1GB$) contains $16M$ records}.
Both column-transform and aggregate operations mostly comprise row-wise operations performing serial bit-by-bit data movement between crossbar rows, showing that even the bulk-bitwise logic latency is chiefly dedicated to data movements. This result is due to the crossbar row-wise operation restriction, requiring that only one column be active at a time. 
We analyze the case where row-wise operations are allowed to operate on multiple columns in any combination (only increasing the row-wise data movement bandwidth). Our analysis shows that the full queries' bulk-bitwise logic latency can be reduced by $80\%$--$86\%$, and the execution time improved by \revMark{$25\%$ for} \myemph{\revMark{Q1}} \revMark{and} \myemph{\revMark{Q6}} \revMark{and by $39\%$} for \myemph{\revMark{Q22\_sub}}, turning the bulk-bitwise logic latency from \revMark{$29\%$--$48\%$} into \revMark{$5.4\%$--$15\%$} of execution time.

Our results show that despite using a non-optimum loop-based n-bit operation design and restricted crossbar operations,
the bulk-bitwise PIM latency used for computation, as opposed to data movement between crossbar rows, does not have a major impact on query execution time (less than 1\% for filter-only queries and less than \revMark{6\%} for full queries \revMark{on large relations}). These results, although dependent on data set size, signal that bulk-bitwise PIM computation latency can be traded-off with other system aspects, \textit{e.g.}, complexity, area, and power.

\subsection{Area}
\label{subsubsec:PIM_module_evaluation}

To estimate the area of the PIM controller, we implemented it in Verilog, synthesized it, and evaluated its area using Cadence Innovus and Synopsys Design Compiler with TSMC 28nm technology. For the PIM module chip area, NVSim~\cite{NVSim} was modified to include a single PIM controller per 64 subarrays. Fig.~\ref{fig:area} shows a PIM module chip area breakdown, with the PIM controller consuming only 0.17\% of chip area.

\subsection{Power and Energy}
To evaluate the system's energy and power consumption, we evaluate and sum the energies of the DRAM main memory, the host, and the PIM modules. The DRAM energy is evaluated by the integrated gem5 DRAM power model~\cite{Chandrasekar2011}. 
The McPAT~\cite{mcpat} tool is used to assess the host power consumption.

\begin{figure}[t]
\centering
\subfloat[Filter-only queries]{\includegraphics[width=0.60\linewidth]{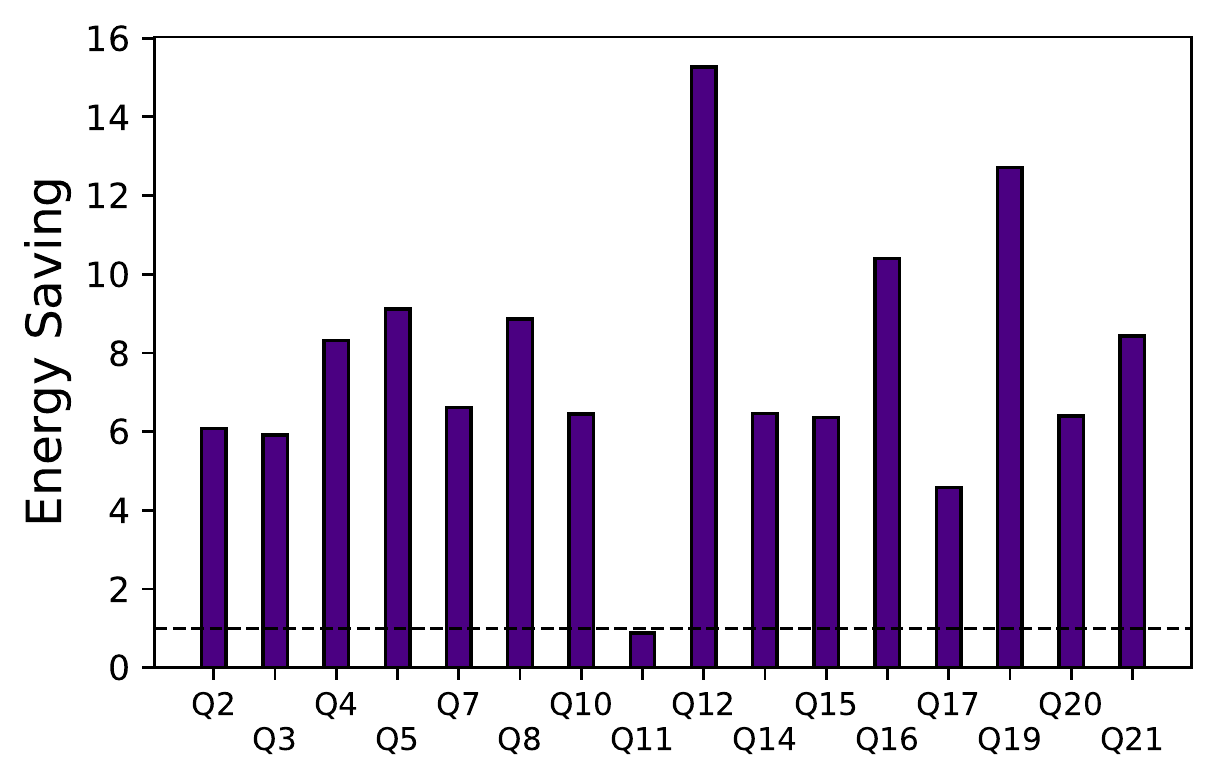}}\hspace{5pt}%
\subfloat[Full queries\\\revMark{(filter+aggregation)}]{\includegraphics[width=0.27\linewidth,trim=-7pt 0pt -7pt 0pt,clip]{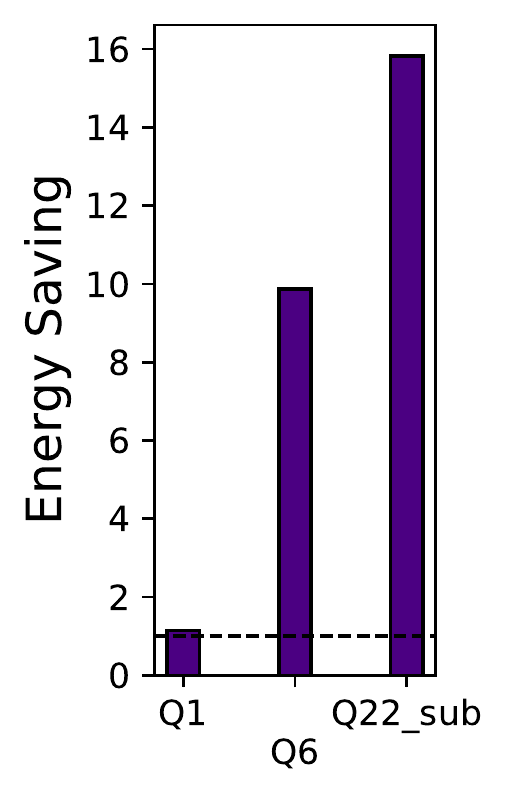}}%
\caption{PIMDB energy saving over baseline.}%
\label{fig:energy}%
\end{figure}

\begin{figure}[t]
\centering
\subfloat[Filter-only queries]{\includegraphics[width=\fullwidt\linewidth]{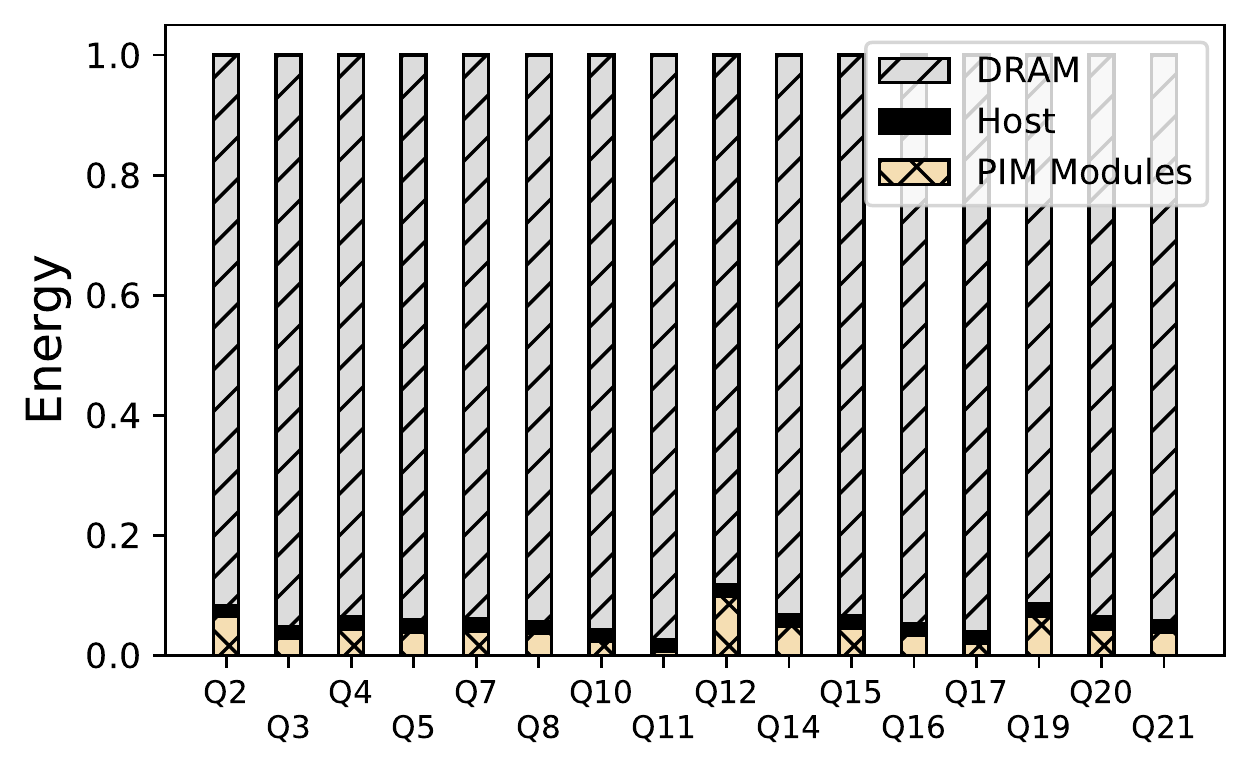}}\hspace{5pt}%
\subfloat[Full queries\\\revMark{(filter+aggregation)}]{\includegraphics[width=\filterwidt\linewidth,trim=-7pt 0pt -7pt 0pt,clip ]{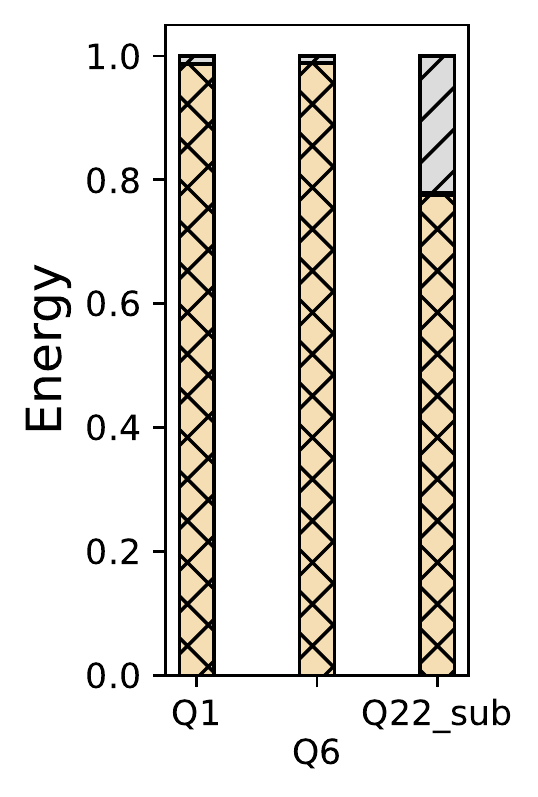}}%
\caption{PIMDB energy breakdown.}%
\label{fig:energy_brekdown_total}%
\end{figure}

The PIM modules' energy is taken as the sum of the PIM controller, bulk-bitwise (stateful) logic, read and write operations, and chip IO energies according to the simulation behavior. The PIM controller energy is evaluated using the Synopsys Design Compiler as explained in Section~\ref{subsubsec:PIM_module_evaluation}. The energy per stateful logic operation was taken from~\cite{Talati2016}, and the energies per read and write operations were taken from~\cite{CONCEPT}. Table~\ref{table:arch_details} lists the different energy values. For chip IO energy costs, the gem5 DRAM energy model was used.
The power of a PIM module is sampled as the average power in a $100ns$ time window.

\begin{figure}[t]
\centering
\subfloat[Filter-only queries]{\includegraphics[width=\fullwidt\linewidth]{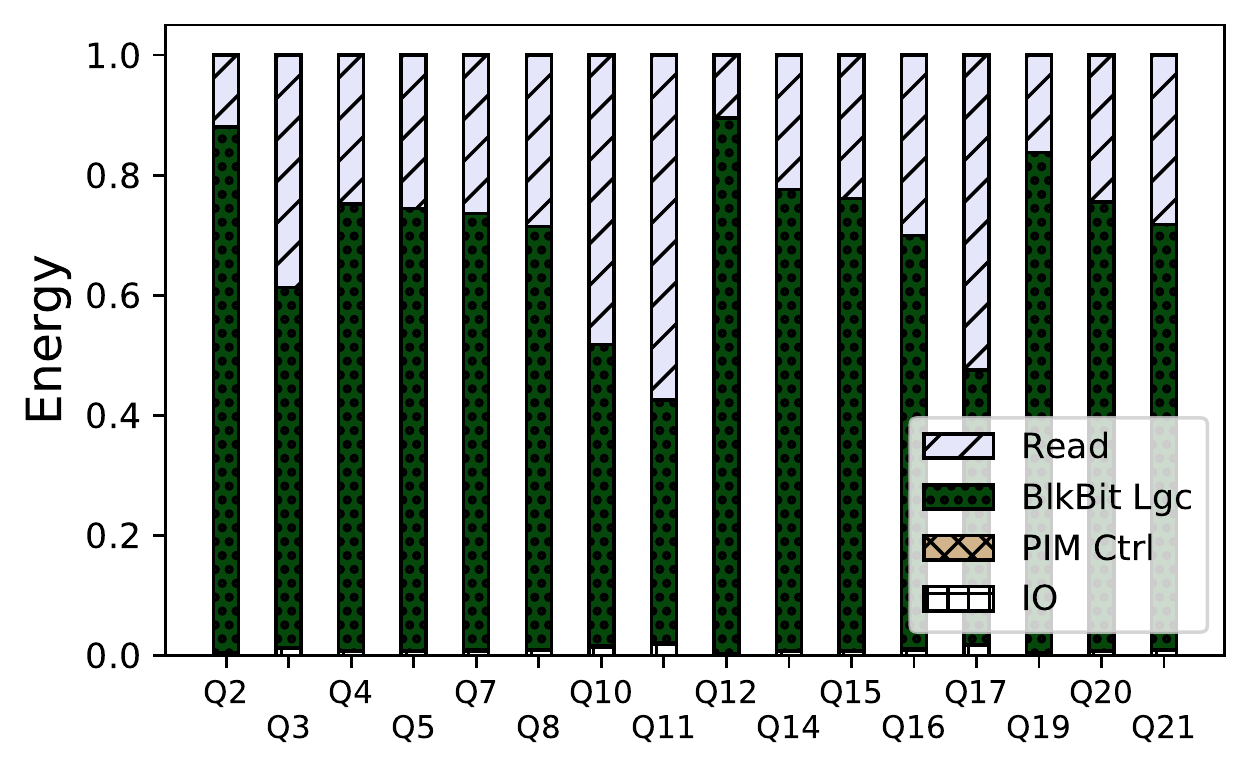}}\hspace{5pt}%
\subfloat[Full queries \\\revMark{(filter+aggregation)}]{\includegraphics[width=\filterwidt\linewidth,trim=-7pt 0pt -7pt 0pt,clip ]{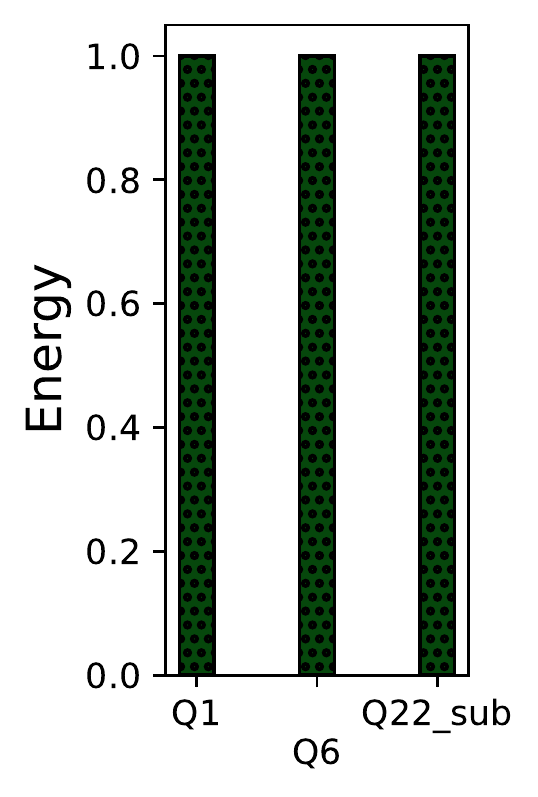}}%
\caption{PIM module energy breakdown.}%
\label{fig:energy_brekdown_PIM}%
\end{figure}

\begin{figure}[t]
\centering
\subfloat[Filter-only queries]{\includegraphics[width=\fullwidt\linewidth]{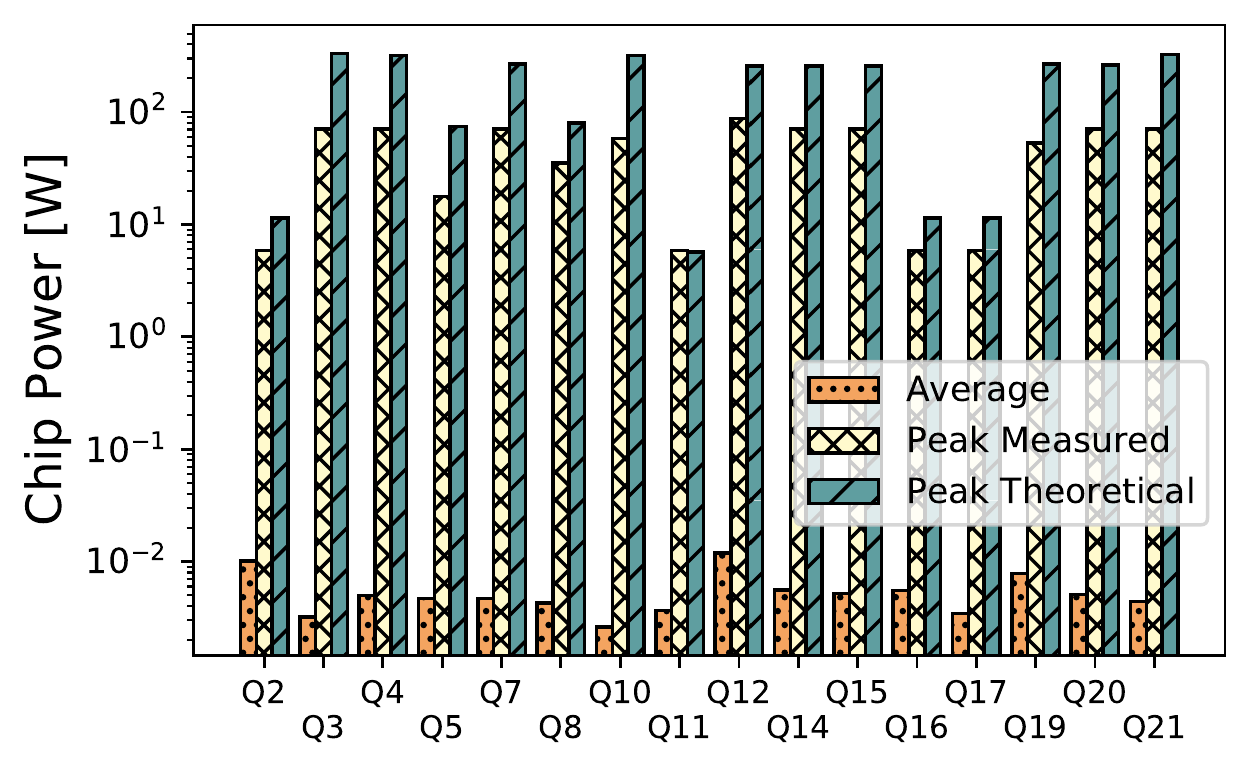}}\hspace{5pt}%
\subfloat[Full queries\\\revMark{(filter+aggregation)}]{\includegraphics[width=\filterwidt\linewidth,trim=-7pt 0pt -7pt 0pt,clip ]{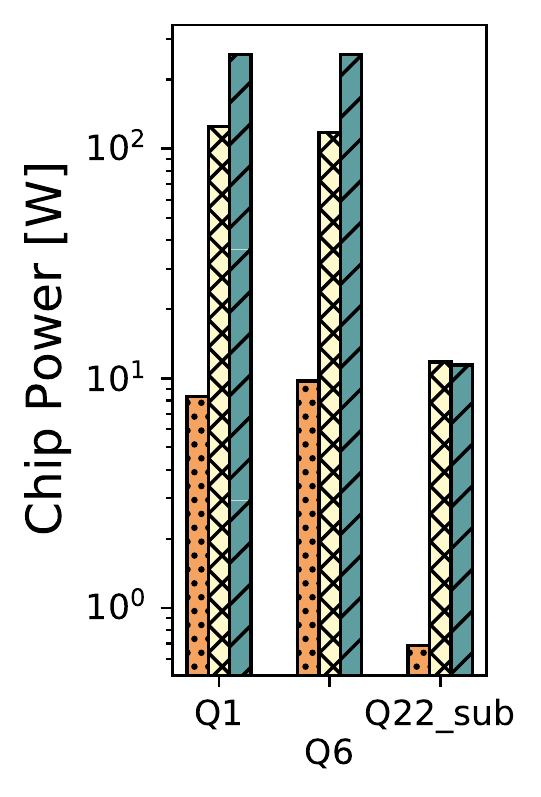}}%
\caption{Peak and average power demand for PIM module chip.}%
\label{fig:power}%
\end{figure}

The energy ratio between the baseline (host and DRAM) and PIMDB (host, DRAM, and PIM modules) is shown in Fig.~\ref{fig:energy}. Overall, an energy reduction of \revMark{$0.88\times$--$15.3\times$ }is achieved for filter-only queries, and a reduction of \revMark{$1.14\times$ and $15.8\times$} is achieved for the full queries.
\myemph{Q1} \revMark{has a relatively small energy reduction} as it performs many reduction operations, completely offsetting the memory traffic energy saving.
The energy breakdowns of the PIMDB system and the PIM module are shown in Figs.~\ref{fig:energy_brekdown_total} and \ref{fig:energy_brekdown_PIM}, respectively.
For the filter-only queries, most of the energy is consumed by the DRAM module (standby energy). While in the PIM modules, most of the energy is dedicated to bulk-bitwise (stateful) logic operations. For the full queries, the PIM modules dominate the energy consumption. For \myemph{Q22\_sub}, the PIM energy component is more moderate than in \myemph{Q1} and \myemph{Q6} since \myemph{Q22\_sub} operates on a smaller relation (\myemph{CUSTOMER} vs. \myemph{LINEITEM}). At the PIM module, more than $99\%$ of the energy is spent on bulk-bitwise logic for the full queries. This is mainly due to the substantial reduction in read operations and longer PIM operations of full queries. Our results indicate that 
the main energy-consuming component is now the bulk-bitwise logic.
In all queries, the host uses relatively little energy since, using PIM, the host is mainly involved in producing the memory operations (read, write, and PIM request), which require light arithmetic. As our workload was chosen to be memory-bound, this behavior is expected.

Fig.~\ref{fig:power} shows the peak and average chip power demand measured by simulations, as well as theoretical peak chip power demand. The theoretical peak power demand is computed as the power required to perform a stateful logic operation across all accessed pages, for the PIM module with the maximum number of pages. The theoretical peak power demand shows that when all pages accessed by a query are operating in parallel, the power demand can reach up to $330W$ per chip, which is high but reasonable for such an accelerator~\cite{TeslaP100}. The peak and average measured power, however, is substantially lower, up to $125W$ and \revMark{$10W$}, respectively, since pages are not synchronized in their operations and stateful logic operates only in part of the query execution. If we look at a bulk-bitwise stateful logic operation across all crossbars in a PIM module chip, which no query in the evaluation performs, we see that the chip power demand can reach $730W$. These results indicate that power-aware scheduling for the PIM operations is required.


\begin{figure}[t]
\centering
\subfloat[Filter-only queries]{\includegraphics[width=\fullwidt\linewidth]{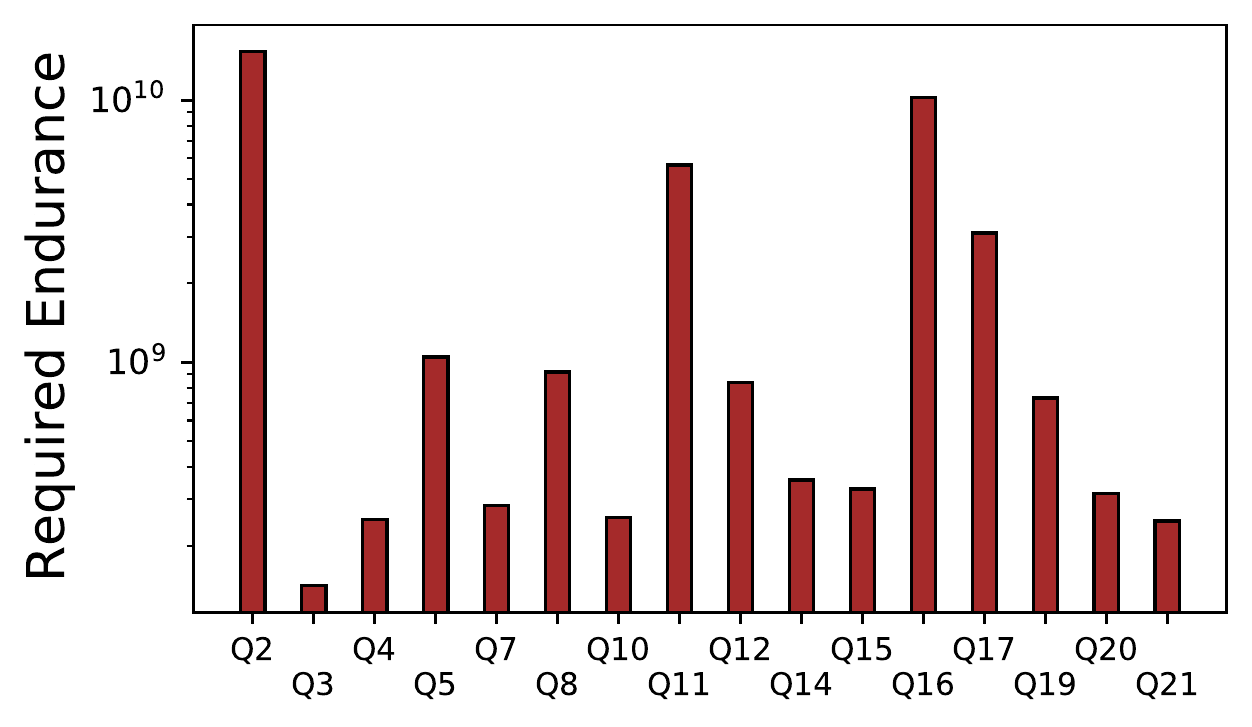}}\hspace{5pt}%
\subfloat[Full queries\\\revMark{(filter+aggregation)}]{\includegraphics[width=0.27\linewidth]{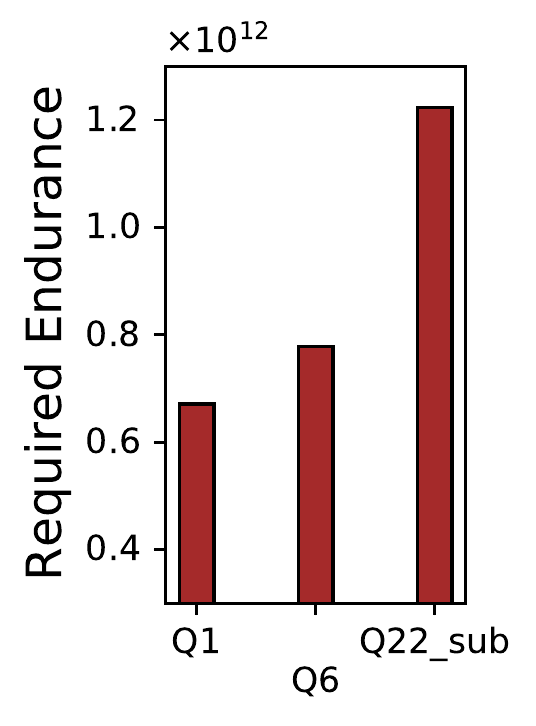}}%
\caption{Required endurance for  a ten-year operation with a 100\% duty cycle.} 
\label{fig:endurance}
\end{figure}

\begin{table}[t]
\centering
\setlength{\tabcolsep}{1.5pt}
\begin{tabular}{|c|c|c|G|I||c|c|c|G|I|} 
\hline
\rule{0pt}{1.5ex}&\bf{Filter}&\bf{Arith.}&\bf{Col. trans.}&\bf{Agg. col/row}&&\bf{Filter}&\bf{Arith.}&\bf{Col. trans.}&\bf{Agg. row/col}\\\hline
\rule{0pt}{1.4ex}\myemph{Q2}&91\%&0&9\%&0/0&\myemph{Q3}&60\%&0&40\%&0/0\\\hline
\rule{0pt}{1.4ex}\myemph{Q4}&77\%&0&23\%&0/0&\myemph{Q5}&77\%&0&23\%&0/0\\\hline
\rule{0pt}{1.4ex}\myemph{Q7}&76\%&0&24\%&0/0&\myemph{Q8}&76\%&0&24\%&0/0\\\hline
\rule{0pt}{1.4ex}\myemph{Q10}&77\%&0&23\%&0/0&\myemph{Q11}&26\%&0&74\%&0/0\\\hline
\rule{0pt}{1.4ex}\myemph{Q12}&91\%&0&9\%&0/0&\myemph{Q14}&80\%&0&20\%&0/0\\\hline
\rule{0pt}{1.4ex}\myemph{Q15}&78\%&0&22\%&0/0&\myemph{Q16}&81\%&0&19\%&0/0\\\hline
\rule{0pt}{1.4ex}\myemph{Q17}&37\%&0&63\%&0/0&\myemph{Q19}&90\%&0&10\%&0/0\\\hline
\rule{0pt}{1.4ex}\myemph{Q20}&77\%&0&23\%&0/0&\myemph{Q21}&77\%&0&23\%&0/0\\\hline
\rule{0pt}{1.4ex}\myemph{Q1}&1\%$>$&8\%&0&85\%/ 7\%&\myemph{Q6}&2\%&23\%&0&68\%/ 6\%\\\hline
\rule{0pt}{1.4ex}{\scriptsize$\myemph{Q22\_sub}$}&6\%&1\%&0&87\%/ 6\%&&&&&\\\hline
\end{tabular}
\vspace{3pt}
\caption{Breakdown of PIM operations' contribution to endurance requirements. The Agg. col/row column shows the portion of the aggregation column and row operations, respectively.}
\label{table:endurance_breakdown}
\end{table}

\subsection{Endurance}
\label{subsec:endurance}
As device endurance is a main challenge for memristive technologies~\cite{Zahoor2020}
, the required endurance for the proposed design and workloads are evaluated. 
For this evaluation, we assume that through the life of the PIM module, the computation 
at a crossbar row is uniformly distributed across all cells of that row. This is a reasonable assumption since the locations of all values in a crossbar row are controlled by software and can be shifted periodically. Furthermore, previously proposed techniques~\cite{N3XT} can be applied to achieve a uniform distribution. Under this assumption, the maximum number of operations on a cell is calculated as follows: the maximum number of operations experienced by a single crossbar row is extracted per query and divided by the number of crossbar row cells. Note that not all crossbar rows experience the same number of operations. Different relations experience different operations and the column-transform and reduce operations do not operate on all crossbar rows uniformly.
Fig.~\ref{fig:endurance} shows the maximum number of operations applied per cell, for each query, where our measured execution (Section~\ref{subsec:query_execution}) is performed back-to-back, \emph{i.e.}, with a 100\% duty cycle, for ten years. Based on previously reported RRAM endurance numbers of $10^{12}$ cycles~\cite{Zahoor2020}
, the results show that the PIMDB lifetime can exceed ten years\revMark{, with the exception of }\myemph{\revMark{Q22\_sub}}\revMark{ due to its operation on a relatively small relation, resulting in frequent writes to the same memory cells.}

Table~\ref{table:endurance_breakdown} shows the breakdown of operations contributing to the maximum number of operations per cell. For the filter-only queries, the filter PIM operations are the dominant contributors (as opposed to the PIM operation latency). This is because the column-transform operation mainly performs bit-by-bit movements between rows, resulting in few operations per row. For the full queries, the reduce operations are still the dominant factor but with the column-wise operations rather than the row-wise operations. These results imply that endurance, rather than latency, may be more important for arithmetic design with stateful logic.

Overall, we see that there are two reasons for the low number of operations applied per cell. First, most of the PIMDB execution time is spent on read operations and not PIM operations, as shown in Fig.~\ref{fig:runtime_breakdown}. Second, the most time-consuming PIM operations have a low number of operations per cell. These operations are the column-transform and row-wise operations for aggregation, for the filter-only and full queries, respectively.

\section{Related Work}
\label{sec:related_work}

Several bulk-bitwise PIM technologies and architectures were previously proposed~\cite{AMBIT,CONCEPT,Giannopoulos2020,Pinatubo,SIMDRAM,RACER}. These works showed how to perform the bulk-bitwise PIM operations, but did not delve into full system performance. Furthermore, these works only demonstrated bulk-bitwise PIM on micro-benchmarks (including database micro-benchmarks), missing the implementation details and operation mix of a standard benchmark.
All these works only show bulk-bitwise operations on either crossbar bitlines or wordlines, enabling only the query's filter, not aggregation, to be performed with solely bulk-bitwise PIM. \revMark{As such, our work has a more comprehensive and more fundamental evaluation of bulk-bitwise PIM than previous works in general, and on database applications specifically.} 
As the evaluation of the filter-only queries shows, the read latency, and not the PIM latency, dominates the query latency.
Hence, other technologies will have similar query execution behavior as PIMDB\revMark{, with execution time differences depending on the difference in read latency from the PIM memory}. The work in \cite{RACER} also supports direct inter-crossbar communication
in addition to bulk-bitwise PIM. This communication enables aggregation without the host, potentially increasing performance further with additional hardware and control costs.

An RRAM-based CAM architecture for database acceleration was presented in~\cite{NVQuery}. The design included analog and stateful logic techniques on the same CAM crossbars, making it hard to distinguish the contribution of each technique separately and involving further hardware and control overheads. While PIMDB and~\cite{NVQuery} support similar basic database primitives, they significantly differ in hardware, operation implementations, relation layout, and host--PIM interaction. 
Moreover, \cite{NVQuery} did not include a programming model and their evaluation consists of only a few custom micro-benchmarks, rather than a standard benchmark.

To the best of our knowledge, none of the previous works for bulk-bitwise operations, where the PIM module is part of the main memory, presents a full programming model~\cite{FloatPIM,NVQuery,CONCEPT,SIMDRAM,RACER}. \cite{Zhou2021} presented a high-level programming language interface for bulk-bitwise PIM. \cite{SIMDRAM} and~\cite{Pinatubo} presented a programming model that avoids the PIM memory management and addressing issues, delegating them to the operating system (OS) to handle. 
These solutions are complementary to our programming model as they are on different abstraction levels. \cite{SIMDRAM} also suggested a host instruction set extension to issue PIM instructions similar to our PIM requests. \cite{Ahn2015} and~\cite{Nai2017} proposed instruction offloading schemes for near-memory architectures. 
Near-memory uses distinct components for processing elements and memory, having different characteristics than bulk-bitwise PIM. \cite{Ahn2015} proposed a host instruction set extension to indicate PIM-enabled instructions, similar to the PIM request proposed in our work, while~\cite{Nai2017} suggested intercepting the host processor atomic instructions for PIM execution.

In-storage~\cite{Xu2020,FlexPushdownDB} and near-memory~\cite{Kepe2019} are other processing techniques that are similar to bulk-bitwise PIM. Both in-storage and near-memory processing try to reduce data movement by bringing the computation closer to where the data reside. Both techniques, however, use standard CMOS technology (\emph{e.g.}, cores, dedicated accelerators), rather than the data storage medium, as the processing elements. Hence, although in-storage and near-memory processing goals are similar to those of bulk-bitwise PIM, the control and operation are different. Additionally, commercial in-storage processing~\cite{FlexPushdownDB} offers to accelerate filtering and aggregation for databases, which are the same primitives inherently available by bulk-bitwise PIM. As mentioned, the approach to accelerating these primitives differs from bulk-bitwise PIM, producing different characteristics and trade-offs.
\revMark{ Nevertheless, in-storage and near-memory processing are not competing with bulk-bitwise PIM, but complementing it. As in-storage and near-memory use logic outside of the memory array, they can be used as hosts for bulk-bitwise PIM, benefiting from the data movement reduction achieved by bulk-bitwise PIM.
}

\section{Conclusion}
In this paper, we investigated bulk-bitwise PIM by proposing PIMDB, a bulk-bitwise PIM architecture based on memristive stateful logic for accelerating analytical processing operations on relational databases, including a host programming model. By focusing on bulk-bitwise PIM for a real-world application and introducing a programming model, the properties of this PIM method were demonstrated. Insights about application mapping, execution and energy breakdowns, PIM logic latency, endurance, and PIM module power were offered. The design reduced the data transfer in the system by supporting filtering and aggregation operations within the memory and included mapping and logic techniques to support these operations. 
The programming model, as well as other aspects of our work, can be applied to other bulk-bitwise PIM technologies and architectures.
PIMDB accelerates TPC-H filter operations by 1.6$\times$--18$\times$ and full queries by 56$\times$--608$\times$, while achieving a 1.7$\times$--18.6$\times$ and 0.81$\times$--12$\times$ energy reduction for these benchmarks, respectively.


%





\ifCLASSOPTIONcaptionsoff
  \newpage
\fi


\bibliographystyle{IEEEtranS}
\bibliography{ref}

%

%

\begin{IEEEbiography}[{\includegraphics[width=1in,height=1.25in,clip,keepaspectratio]{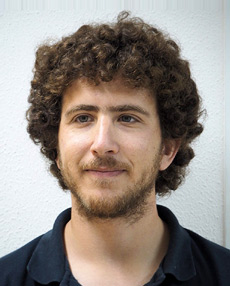}}]{Ben Perach}
is a Ph.D. candidate at the Andrew and Erna Viterbi Faculty of Electrical and Computer Engineering, Technion – Israel Institute of Technology. Ben received his B.Sc. degree in mathematics from The Hebrew University of Jerusalem in 2010 and his M.Sc. degree in electrical engineering from Tel Aviv University in 2017. Ben's current research interests include computer architecture with a focus on processor design, and also field-programmable gate arrays, security, and data networks.
\end{IEEEbiography}

\begin{IEEEbiography}[{\includegraphics[width=1in,height=1.25in,clip,keepaspectratio]{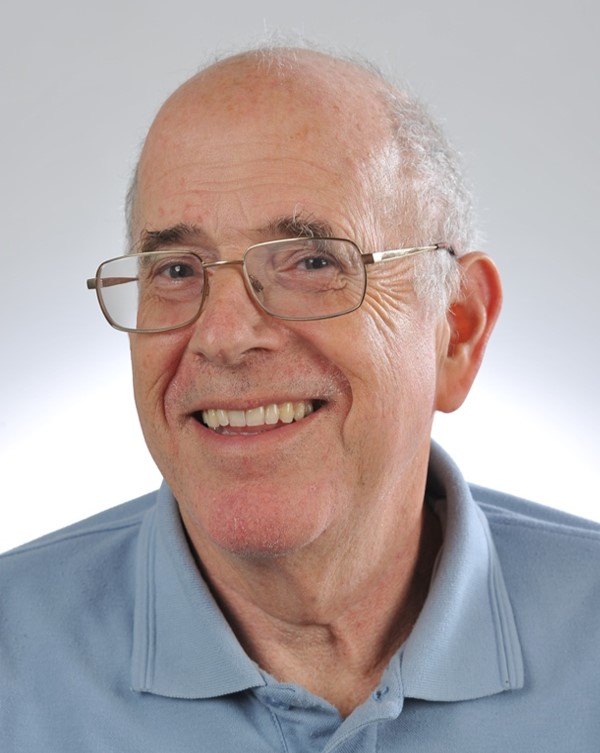}}]{Ronny Ronen}(Fellow, IEEE)
received his B.Sc. and M.Sc. degrees in computer science from the Technion, Haifa, Israel, in 1978 and 1979, respectively. He is a Senior Researcher at the Andrew and Erna Viterbi Faculty of Electrical \& Computer Engineering at the Technion. He was with Intel Corporation from 1980 to 2017 in various technical and managerial positions. In his last role there, he led the Intel Collaborative Research Institute for Computational Intelligence. He was the Director of Microarchitecture Research and a Senior Staff Computer Architect at the Intel Haifa Development Center until 2011. He led the development of several system software products and tools, including the Intel Pentium processor performance simulator and several compiler efforts. In these roles, he led/was involved in the initial definition and pathfinding of major leading-edge Intel processors. He holds over 80 issued patents and has published over 35 papers.
\end{IEEEbiography}

\begin{IEEEbiography}[{\includegraphics[width=1in,height=1.25in,clip,keepaspectratio]{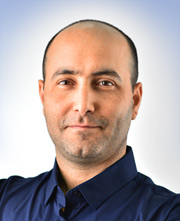}}]{Benny Kimelfeld}
is a Professor at the Computer Science Faculty at Technion, Israel, and the head and founder of the Technion Data and Knowledge (TD\&K) Laboratory. After receiving his Ph.D. from The Hebrew University of Jerusalem, and before joining Technion, he has been a Research Staff Member at IBM Research Almaden, then a Computer Scientist at LogicBlox. Benny’s research spans a spectrum of both foundational and systems aspects of data management, such as probabilistic and inconsistent databases, information retrieval over structured data, and infrastructure for text analytics. Benny was the program-committee chair of the 2018 International Conference on Database Theory (ICDT), a co-chair of the 2016 Web and Databases Workshop (WebDB), and a co-chair of the 2014 SIGMOD/PODS workshop on Big Uncertain Data (BUDA). He currently serves as an associate editor in the Journal of Computer and System Sciences (JCSS).
\end{IEEEbiography}


\begin{IEEEbiography}[{\includegraphics[width=1in,height=1.25in,clip,keepaspectratio]{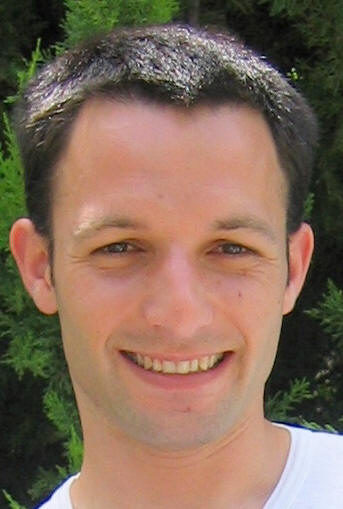}}]{Shahar Kvatinsky}(Senior Member, IEEE) is an Associate Professor at the Viterbi Faculty of Electrical and Computer Engineering, Technion. Shahar received the B.Sc. degree in Computer Engineering and Applied Physics and an MBA degree in 2009 and 2010, respectively, both from the Hebrew University of Jerusalem, and the Ph.D. degree in Electrical Engineering from the Technion in 2014. From 2006 to 2009, he worked as a circuit designer at Intel. From 2014 to 2015, he was a post-doctoral research fellow at Stanford University. Kvatinsky is a member of the Israel Young Academy. He is the head of the Architecture and Circuits Research Center at the Technion and chair of the IEEE Circuits and Systems in Israel. Kvatinsky has been the recipient of numerous awards including: 2020 MDPI Electronics Young Investigator Award, 2019 Wolf Foundation's Krill Prize, 2015 IEEE Guillemin-Cauer Best Paper Award, ERC starting grant, and the 2017 Pazy Memorial Award.
\end{IEEEbiography}




\end{document}